\documentclass{aa}
\usepackage[varg]{txfonts}
\usepackage{graphicx,textcomp}
\usepackage{booktabs}
\usepackage{array}
\usepackage{tablefootnote}

\usepackage[breaklinks=true, colorlinks=true, linkcolor=blue, citecolor=blue]{hyperref}
\usepackage{url}
\usepackage{float}
\usepackage{threeparttable}
\usepackage{stfloats}
\usepackage{multirow}
\usepackage{rotating}
\usepackage{arydshln}
\usepackage[all]{hypcap}
\usepackage[capitalise]{cleveref}
\usepackage[separate-uncertainty]{siunitx}
\usepackage{natbib}
\bibpunct{(}{)}{;}{a}{}{,} 
\usepackage{txfonts}
\usepackage{xcolor}
\usepackage{ulem}
\usepackage{textcomp}

\newcommand{\ero}{eROSITA}

\newcommand{\rfive}{$R_{500}$}
\newcommand{\rtwo}{$R_{200}$}

\let\unit\si

\begin{document}

\title{The SRG/eROSITA All-Sky Survey}
\subtitle{A comprehensive X-ray analysis of the Hydra I galaxy cluster}

\author{Alpish Srivastava\inst{1,2,3,}\thanks{Corresponding author: \href{mailto:alpish.srivastava@roma2.infn.it}{{alpish.srivastava@roma2.infn.it}}}
     \and Thomas H. Reiprich\inst{1}
     \and Angie Veronica\inst{1}
     \and Florian Pacaud \inst{1} 
     \and Jakob Dietl \inst{1,4}
     \and Fiona Knies \inst{5} 
     \and Manami Sasaki \inst{5}
    }
\authorrunning{A. Srivastava et al.}
\titlerunning{A comprehensive X-ray analysis of the Hydra I galaxy cluster}

\institute{Argelander-Institut für Astronomie (AIfA), Universität Bonn, Auf dem Hügel 71, 53121 Bonn, Germany
  \and Dipartimento di Fisica, Università degli studi di Roma ‘Tor Vergata’, Via della Ricerca Scientifica, 1, 00133 Roma, Italy
  \and INFN, Sezione di Roma 2, Università degli studi di Roma ‘Tor Vergata’, Via della Ricerca Scientifica, 1, 00133 Roma, Italy
  \and Max-Planck-Institut für Radioastronomie, Auf dem Hügel 69, 53121 Bonn, Germany
  \and Dr. Karl Remeis Observatory and ECAP, Universität Erlangen-Nürnberg, Sternwartstraße 7, 96049 Bamberg, Germany
 }

\date{Received 03 December 2025 / Accepted 01 June 2026}

\abstract 
{The Hydra I galaxy cluster (Abell 1060) is a nearby example of a low-temperature cluster that exhibits intermediate cool core (CC) and non-cool core (NCC) properties. In addition, little is known about the gas properties and large-scale structure beyond its \rfive.} 
{Our aims are to extend the characterization of the intracluster medium (ICM) properties at least until \rtwo, and to study the correlation between the X-ray emission and nonthermal emission within $R=0.15$\rfive, and the optical/IR galaxy distribution beyond \rtwo.} 
{We used data from the first four SRG/eROSITA All-Sky Surveys (eRASS:4) and an archival \textit{Chandra} observation to image the X-ray emission from Abell 1060. We also used multiwavelength data from TGSS (radio), 2MASS (IR), and NED (optical) to investigate the nonthermal emission, 2D galaxy distribution, and its redshift evolution, respectively. The surface brightness and spectral analyses are also extended until 3\rtwo\;and \rtwo, respectively, following a detailed cosmic X-ray background (CXB) analysis.} 
{Our fully corrected eROSITA image reveals a relaxed ICM morphology within \rfive. We detected two surface brightness discontinuities near the central galaxy NGC~3311 that spatially coincide with diffuse radio emission along the line of sight. We also modeled the central surface brightness cusp and the full profile until 3\rtwo\;using a modified $\beta$-model. Furthermore, we detected two soft X-ray excesses with high spatial correlation with the 2D optical galaxy distribution beyond \rtwo. In particular, the excess in the north has a significance of 3.9$\sigma$ above the local CXB level. This suggests that the outskirts of Abell 1060 are actively accreting baryons. The NED spectroscopic redshift distribution of member galaxies is unimodal with a best-fit mean and standard deviation of $0.0121\pm0.0027$ from a Gaussian fit. We also estimate the average ICM temperature and metallicity of $\langle k_\mathrm{B}T \rangle=2.51\substack{+0.21\\-0.21}\thinspace\unit{keV}$ and $\langle Z\rangle=0.19\substack{+0.10\\-0.10}\thinspace Z_\odot$, respectively, from the 0.2--0.5\rfive\;annulus. Overall, the temperature profile is broadly consistent with the average temperature profiles from hydrodynamical simulation and \textit{Suzaku} between 0.43\rfive\;and \rfive. \\
}
{}

\maketitle
\nolinenumbers

\section{Introduction}
\label{sec:intro}
Galaxy clusters are the largest gravitationally bound and virialized objects that originate from the growth and subsequent collapse of initial density perturbations in the early Universe. They serve as large-scale hydrodynamic laboratories that allow us to probe the dynamical state of clusters through X-ray observations of the intracluster medium (ICM). Traditionally, clusters have been classified as either cool core (CC) or non-cool core (NCC) based on the presence of an inward declining ICM temperature structure within $R\sim0.2$\rtwo\;\citep[e.g.,][]{2001ApJ...560..194M}. Subsequently, \citet{2010A&A...513A..37H} subclassified CCs into strong cool cores and weak cool cores (WCCs) based on their central cooling time, entropy, and the relative position between the X-ray peak and the brightest cluster galaxy. One such WCC candidate is the Hydra I cluster (Abell 1060), which is a member of the Hydra-Centaurus supercluster \citep[e.g.,][]{2001hydcentdist,largehc}. Due to its proximity (\cref{tab:clusterparams}), it is highly extended in the sky with an \rfive~=~$\ang[angle-symbol-over-decimal]{;45.38}$ \citep[][]{mcxc}. The first observations of the cluster and its globular cluster distribution were performed in the optical and radio regimes \citep[e.g.,][]{1966hydra,1976hydra,1980MNRAS.190..631S,1983A&A...125..187R}. Specifically, \citet{1982hydra} reported a recession velocity of $v_0=3425\pm34$\thinspace\unit{\kilo\metre\per\second} and a relatively low line of sight velocity dispersion of 676\thinspace\unit{\kilo\metre\per\second}. \\
\indent
The inner $R\approx0.4$\rfive\;of the cluster has been extensively studied in X-rays. Early observations with the EXOSAT and ASCA X-ray telescopes \citep[e.g.,][]{1988hydra,1996ascahydra,2001ascahydra} revealed that the X-ray emission in the core is dominated by the two central galaxies, NGC~3311 and NGC~3309, and the cluster appears to be relaxed. They further reported a total mass (calculated assuming hydrostatic equilibrium) of $M_\mathrm{tot}\approx10^{14}\thinspace M_\odot$ and flat temperature and metallicity profiles within $R=\ang{;20}$, with average values of $k_\mathrm{B}T\approx3.2$\thinspace\unit{keV} and $Z\approx0.3\thinspace Z_\odot$, respectively. Similar results were also obtained by \citet{2000hydra} using ROSAT PSPC data. In the next decade, several studies utilized \textit{Chandra}'s superior angular resolution and \textit{XMM-Newton}'s large effective area to study the X-ray emission from the ICM in detail. \citet{2002hydra} reported that the central galaxies were noted to have small extents (<\ang{;;20}), but showed no signs of mutual stripping. They also confirmed that the ICM exhibited a relaxed morphology. Later, \mbox{\citet{2004hydra,2006xmmhydra}} observed a peak in the temperature profile of the previously believed isothermal ICM at $R=\ang[angle-symbol-over-decimal]{;2.25}$, followed by a 30\% decline in the outskirts. They also reported that the concentration of matter in the central surface brightness cusp is localized within the inner $R=\ang[angle-symbol-over-decimal]{;3.23}$. Moreover, they detected the ram-pressure stripped halo of \mbox{NGC 3311}, which yielded a high metallicity of $Z=1.5\thinspace Z_\odot$ and an iron mass of $1.9\times10^{7}\thinspace M_\odot$ within $R\lesssim\ang{;5}$. Last but not least, \citet{2007hydra} used two \textit{Suzaku} pointings to extend their analysis until $R=\ang{;27}$ toward the east and observed a  decline in the temperature profile similar to that found by \mbox{\citet{2004hydra,2006xmmhydra}}. In the central $R=\ang{;5}$, their temperature profile displayed a flat top, which they believed to be an initial phase of a CC. Furthermore, they observed a decline of about 50\% in the abundance profiles of heavier elements such as Si, S, and Fe. \\
 \begin{figure}[H]
   \centering
   \includegraphics[width=0.49\textwidth]{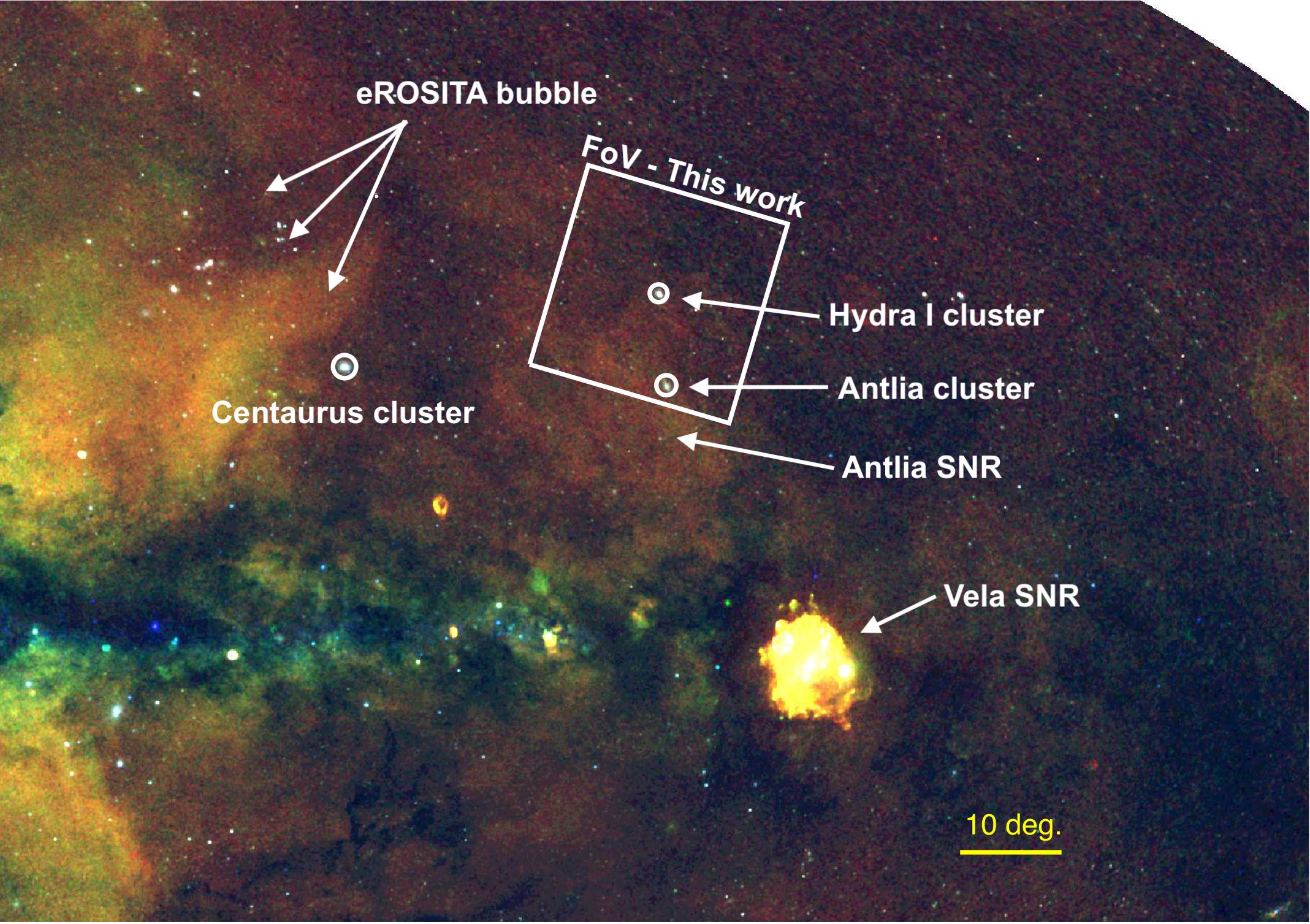}
   \caption[]{Zoomed-in image of half sky eRASS:1 RGB image prepared using broadband TM8 maps in the energy ranges 0.4--0.6\thinspace\unit{keV} (red), 0.6--1.0\thinspace\unit{keV} (green), and 1.0--2.3\thinspace\unit{keV} (blue) from \citet{2024A&A...681A..77Z}. The approximate FoV of our eRASS:4 image and other prominent sources in its vicinity are annotated. The circles on the clusters represent their \rfive.}
   \label{fig:RGBhalfsky}
\end{figure}
 In radio wavelengths, \citet{1985radiohydra} discovered the radio jets of NGC~3309 via the 6 cm continuum observations from the Very Large Array. However, only a weak radio source was detected in NGC 3311. Recently, \citet{2024radiohydra} discovered diffuse radio emission near NCG~3311 using  Giant Metrewave Radio Telescope (GMRT) data, whose northern section coincides with the stripped halo of NGC~3311 along the line of sight. The rest of the emission extends farther to the southeast by approximately \ang{;4}. The source of this emission is unknown, but several plausible scenarios regarding its origin, such as fossil emission from NGC 3311, radio relics, and odd radio circles, are discussed. They also detected radio emission from the spiral galaxy \mbox{NGC 3312}. \\
 \indent
 Additionally, the Antlia cluster is located about \ang{7} south of Abell 1060 (\cref{fig:RGBhalfsky}) and has a slightly lower redshift \mbox{\citep[$z\approx0.01$; e.g.,][]{antliared,2016antlia}}. The large-scale galaxy distribution suggests that a strand of galaxy filament extends from Antlia to the Centaurus cluster, with Abell~1060 possibly associated with this structure \citep[][]{largehc}. Moreover, the highly extended Antlia supernova remnant \citep[SNR;][]{2024A&A...681A..77Z,2026A&A...708A.198K} is also located in projection between the two clusters and dominates the local \mbox{X-ray} foreground. In \cref{fig:RGBhalfsky}, we display the positions of both Abell~1060 and the Antlia cluster relative to the Antlia SNR, the \ero\,bubble \citep[][]{erobub}, and the Galactic plane.\\
 \indent
 The unexplored ICM as well as the large-scale structure beyond \rfive, and the presence of a nonthermal component in the core, make Abell 1060 an ideal candidate to study using \ero\;\citep{erosita} and \textit{Chandra} \mbox{\citep{chandra}}. Recent large field of view (FoV) eROSITA observations of nearby clusters (e.g., Virgo \citep{hannah}, Fornax \citep{fornax}, and Centaurus \citep{angiecent}) have demonstrated its capability to detect ICM emission from well beyond \rtwo. Furthermore, the softer energy response of \ero\;has been critical in the successful detection and analysis of filamentary X-ray emission between and beyond clusters \citep[e.g.,][]{reiprichfila,2024veronica,2024jakob}.\\
 \indent
For this work, we took advantage of the aforementioned properties of eROSITA and \textit{Chandra}'s \ang{;;1} FoV average angular resolution to perform an imaging and spectral analysis of \mbox{Abell 1060} and the surrounding X-ray emission. We focused on the characterization of the surface brightness and gas properties of the ICM at least until \rtwo\;and perform an in-depth analysis of the emission in the inner core. Furthermore, we attempted to correlate the observed X-ray features on all scales with emission in other wavebands in our FoV. The paper is structured  in the following manner. In Sect.\thinspace\ref{sec:datared} we describe the data reduction process and the subsequent analyses. In Sect.\thinspace\ref{sec:results} we describe our results and discuss them. Finally, we summarize our key findings and conclude in Sect.\thinspace\ref{sec:conclusion}. \\ 
\begin{table}
   \centering
   \caption{Cluster parameters of Abell 1060 from the MCXC catalog \citep{mcxc}.}
   \begin{tabular}{@{}cc@{}}
   \toprule
    Parameters  & Value              \\ \midrule
   $z$ & 0.0126    \\
   R.A. [°] &$159.174$  \\
   Dec. [°] &$-27.524$  \\
   $L_{\text{X,500}}$ [erg s$^{-1}$] &$3.11\times 10^{43}$ \\
   $M_{\text{500}}$ $[M_{\odot}]$ &$9.94\times 10^{13}$  \\
   $R_{\text{500}}$ [arcmin] &$45.38$ \\
   \bottomrule
   \end{tabular}
   \tablefoot{The central position (J2000) is given by the R.A. and Dec. $L_{\text{X,500}}$ and $M_{\text{500}}$ are the X-ray luminosity in the 0.1--2.4\thinspace\unit{keV} band and the total mass within $R_{\text{500}}$, respectively.}
   \label{tab:clusterparams}
\end{table}
 \indent
For our analysis we used the physical parameters of Abell~1060 from the Meta-Catalog of the compiled properties of X-ray detected Clusters of galaxies \mbox{\citep[MCXC;][]{mcxc}}, which  are listed in \cref{tab:clusterparams}. Using the $R_\mathrm{500}$ from \cref{tab:clusterparams} and the conversion relations between characteristic radii given in \citet{2013reiprich}, we calculate $R_\mathrm{2500}=\ang[angle-symbol-over-decimal]{;19.55}$, $R_\mathrm{200}=\ang[angle-symbol-over-decimal]{;69.82}$, and $3R_\mathrm{200}~=~\ang[angle-symbol-over-decimal]{;209.46}$ of \mbox{Abell 1060}. Throughout this work, we assume a flat $\Lambda$CDM cosmology with $\Omega_\text{m}=0.3$, $\Omega_\Lambda=0.7$, and $H_0=70\thinspace\unit{km.s^{-1}.Mpc^{-1}}$. All error bars are at the 68.3\% confidence interval. Moreover, the physical angular scale at the redshift of 0.0126 is $\ang{;;1}=0.26\thinspace\unit{kpc}$.

\section{Data reduction and analysis}
\label{sec:datared}
\subsection{eROSITA data reduction}
\label{sec:erodatared}
We used data from the first four SRG/eROSITA All-Sky Surveys (eRASS:1-4) for our analysis. In particular, we utilized 23 sky tiles,\footnote{The entire eRASS sky is divided into 4700 \href{https://erosita.mpe.mpg.de/dr1/AllSkySurveyData_dr1/}{sky tiles} of size $\ang[angle-symbol-over-decimal]{3.6}\times \ang[angle-symbol-over-decimal]{3.6}$. } namely 152117, 153120, 154(114, 123), 155(117, 126), 156120, 157(114, 123), 158(111, 117, 126), 159120, 161(111, 114, 123), 162(117, 126), 163120, 164(114, 123), 165117, and 166120, with Abell 1060 being centered on the tile 158117. The total area of the FoV is roughly \mbox{$\ang{18}\times\ang{18}$} at its maximum extent, and it includes Abell 1060's outskirts (3\rtwo\;and beyond) and \rfive\;of the Antlia cluster (\cref{fig:RGBhalfsky}). For the data reduction, we used the eROSITA Science Analysis Software \citep[eSASS;][]{esass} version \verb|eSASSusers_211214| (eSASS4DR1) with the data processing version c020 and HEASoft\footnote{\href{https://heasarc.gsfc.nasa.gov/docs/software/heasoft/}{https://heasarc.gsfc.nasa.gov/docs/software/heasoft/}} version 6.35. In this work, we adopted the standard naming convention for the eROSITA telescope modules (TMs), i.e., TM1+2+3+4+6~=~TM8 (TMs with on-chip filters), TM5+7~=~TM9 (TMs without on-chip filters), and TM8+9~=~TM0. We also note that, due to the lack of on-chip filters, the low-energy TM9 observations ($\lesssim$0.5\thinspace\unit{keV}) are severely affected by an optical light leak issue \citep[][]{erosita} and are therefore removed from our analysis.\\
\indent
We followed the eROSITA data reduction steps as described in \citet{reiprichfila}, \citet{2024veronica}, \citet{hannah}, \citet{2024jakob}, \citet{angiecent}, and \citet{fornax}, which we summarize in brief here. We used the eSASS task \texttt{evtool} with the arguments \texttt{pattern=15} and \texttt{flag=0xe00fff30} to include single, double, triple, and quadruple patterns and to remove bad pixels and the strongly vignetted corners of the CCDs, respectively. In addition, we applied the original good time intervals (GTIs) using the argument \texttt{gti="GTI"}. We further extracted the light curves in the hard band 5--10\thinspace\unit{keV} for each sky tile using the task \texttt{flaregti} and applied a $3\sigma$ threshold to filter out the soft proton flares (SPFs). The GTIs thus acquired are applied to the event lists using \texttt{evtool} with the argument \texttt{GTI="FLAREGTI"}. For imaging, we primarily used the energy bands \mbox{0.2--2.3\thinspace\unit{keV}} and \mbox{0.5--2.3\thinspace{keV}} for TM8 and TM9, respectively. However, despite the differing low energy limits for TM8 and TM9, we label the soft band for TM0 as \mbox{0.2--2.3\thinspace{keV}} hereafter. We also removed the \mbox{1.35--1.6\thinspace\unit{keV}} band from the observation because the 1.4\thinspace\unit{keV} \mbox{Al-K$\alpha$} line was observed to be more prominent in the filter-wheel-closed (FWC) data \citep[][]{angiecent} and could have biased our estimates of the particle induced background \citep[PIB;][]{fwc,yeungfwc}. \\
\indent
Furthermore, we created the TM0 PIB map following Eq.\thinspace1 from \citet{reiprichfila}. We used the FWC ratio of the soft and hard band (6.7--9.0\thinspace\unit{keV}) counts and the hard band counts from the current observation to estimate the PIB counts in the soft band. We then distributed them throughout our FoV using a normalized flat exposure map (generated using the \texttt{expmap} task). Moreover, we also corrected for the variation in the soft X-ray absorption due to the varying total hydrogen column density along the line of sight in our large FoV \citep[][Sect. 2.1.4]{reiprichfila}. For this, we used a cut and reprojected HI4PI \citep{HI4PI} $N_\mathrm{HI}$ map and an $N_\mathrm{H_2}$ map obtained using the Swift\footnote{\href{https://www.swift.ac.uk/analysis/nhtot/}{https://www.swift.ac.uk/analysis/nhtot/}} tool, and combined the two maps following the method described in \citet{2013willy} to create the $N_\mathrm{H,tot}$ map (\cref{fig:nhtotmap}). We report that the median $N_\mathrm{H,tot}$ within \rtwo\;is $5.79\times10^{20}\thinspace\unit{\centi\metre^{-2}}$. We then extracted 1596 $N_\mathrm{H,tot}$ values from the $N_\mathrm{H,tot}$ map and simulated a spectrum for each value in XSPEC \citep{xspec} version 12.12.0 using the \texttt{fakeit} command and the model
\begin{equation}
    \label{eq:sim_model}
    \texttt{apec}_{\texttt{LHB}} + \texttt{TBabs}* (\texttt{apec}_{\texttt{MWH}} + \texttt{powerlaw}) .
\end{equation}
A detailed description of the model components is mentioned in Sect. \ref{sec:spectralanalysis}. The parameter values for all components are taken from \citet{hannah} and are mentioned in \cref{tab:sim_params}. The metallicity and redshift of the components were fixed to 1 $Z_{\odot}$ and 0, respectively. The simulations are performed for both TM8 and TM9 using their respective response files. From these simulations, we obtained count rates corresponding to each $N_\mathrm{H,tot}$ value. The ratio of these count rates and the count rate corresponding to the median $N_\mathrm{H,tot}$ is used to create a correction factor map that is multiplied to the TM0 exposure map. The variation in correction factors for TM8 and TM9 is 32\% and 28\%, respectively. Finally, we used the TM0 PIB map and the relative $N_\mathrm{H,tot}$ corrected TM0 exposure map to create the fully corrected (background subtracted and exposure corrected) TM0 eROSITA image in the 0.2--2.3\thinspace\unit{keV} band (\cref{fig:cxbsetup}). \\
\vspace{-4mm}
\subsection{Chandra data reduction}
We used the \textit{Chandra} observation (ObsID: 2220) of Abell~1060, which was performed on 4th June 2001 in the VFAINT mode (Very Faint Mode). The total exposure time of the observations is roughly 32\thinspace\unit{ks}. For the data reduction, we used the software \textit{Chandra} Interactive Analysis of Observations \cite[CIAO;][]{ciao} version 4.12 and \texttt{CALDB}\footnote{\href{https://heasarc.gsfc.nasa.gov/docs/heasarc/caldb/caldb_intro.html}{https://heasarc.gsfc.nasa.gov/docs/heasarc/caldb/caldb\_intro.html}} version 4.9.3. Abell~1060 is centered on the ACIS-I3 chip in this observation, and thus we only utilized data from this chip, which covers the central \ang{;8}$\times$\ang{;8} region 
of the cluster. Furthermore, we used the \verb|chandra_repro| script to ensure the correct calibrations were used to produce the calibrated, bad pixels removed, and initial GTIs applied level 2 event list. We extracted the light curve of the observation using the CIAO task \texttt{dmextract} with a time binning of 20\thinspace\unit{s}. We then filtered out the SPF using the task \verb|lc_clean|, which performed a $3\sigma$ clipping around the observed mean count rate of 4.86\thinspace\unit{counts.s^{-1}}. After filtering, we obtained a flare-free exposure time of 31.49\thinspace\unit{ks}, which suggests a low SPF contamination in the data. The new GTIs are then applied to the event list using the task \texttt{dmcopy}. \\
\indent
We then used the \texttt{blanksky} and \verb|blanksky_image| scripts to estimate and subtract the background. The \texttt{blanksky} script matches the level 2 event list with a custom background file in the blank sky and \lq\lq stowed\rq\rq\;background datasets present in \texttt{CALDB}. The background file is then reprojected to
match the observations, and a scale factor is calculated for each chip present in the observation \citep[][]{chandrabg}. The \verb|blanksky_image| script then uses the reprojected background file and the scale factors to create a scaled background image. We used the parameter \verb|weight_method=particle| to ensure that the scaling method scales the background by the ratio of the observation counts to the counts in the band \mbox{9--12\thinspace\unit{keV}}. Through this, we obtained a background subtracted image in the \mbox{0.5--2.3\thinspace\unit{keV}} band. Finally, we used the \texttt{fluximage} task to create the fully corrected flux images in the same energy band. 

\subsection{Imaging analysis}
We performed wavelet filtering of the cleaned TM0 image following the implementation described in \citet{xmmpipe} and \citet{wavelet}, to enhance the faint extended emission and point sources in our image. As we are primarily interested in the emission from the ICM, we used \textsc{SExtractor} \citep{sextractor} to detect and catalog all the point sources, which allowed us to mask them from further analyses. Additionally, we repeated the entire data reduction procedure for three narrow bands, \mbox{0.2--0.8\thinspace\unit{keV}}, \mbox{0.8--1.2\thinspace\unit{keV}}, and \mbox{1.2--2.3\thinspace\unit{keV}} to create a wavelet filtered RGB image (\cref{fig:wfrgb}) that allows us to spatially distinguish between the softer and harder energy components of the diffuse X-ray emission. In addition, we used the \textsc{SciPy} \citep{scipy} implementation of the Gaussian gradient magnitude (GGM) filter to enhance the faint gradients of different scales in our image. Several previous studies have used this technique to enhance gradients and study the gas structure from X-ray images \citep[e.g.,][]{2016MNRAS.460.1898S,ggm,2022ApJ...929...37W,hannah,angiecent}. In this work, we specifically used the kernels with $\sigma=8,\,16,\,32,$ and 64 pixels for eROSITA and $\sigma=10$ pixels for \textit{Chandra}. We avoided excising point sources from the eROSITA image because eROSITA's relatively large PSF results in holes that introduce false gradients in the filtered image. Furthermore, we applied the 6-25 pixels unsharp masking to sharpen the small-scale X-ray features in the \textit{Chandra} image \citep[e.g.,][]{2016MNRAS.460.1898S}. All these images are discussed in Sect.\thinspace\ref{sec:xrayimages}. \\

\begin{figure}[H]
    \centering
    \includegraphics[width=1.0\linewidth]{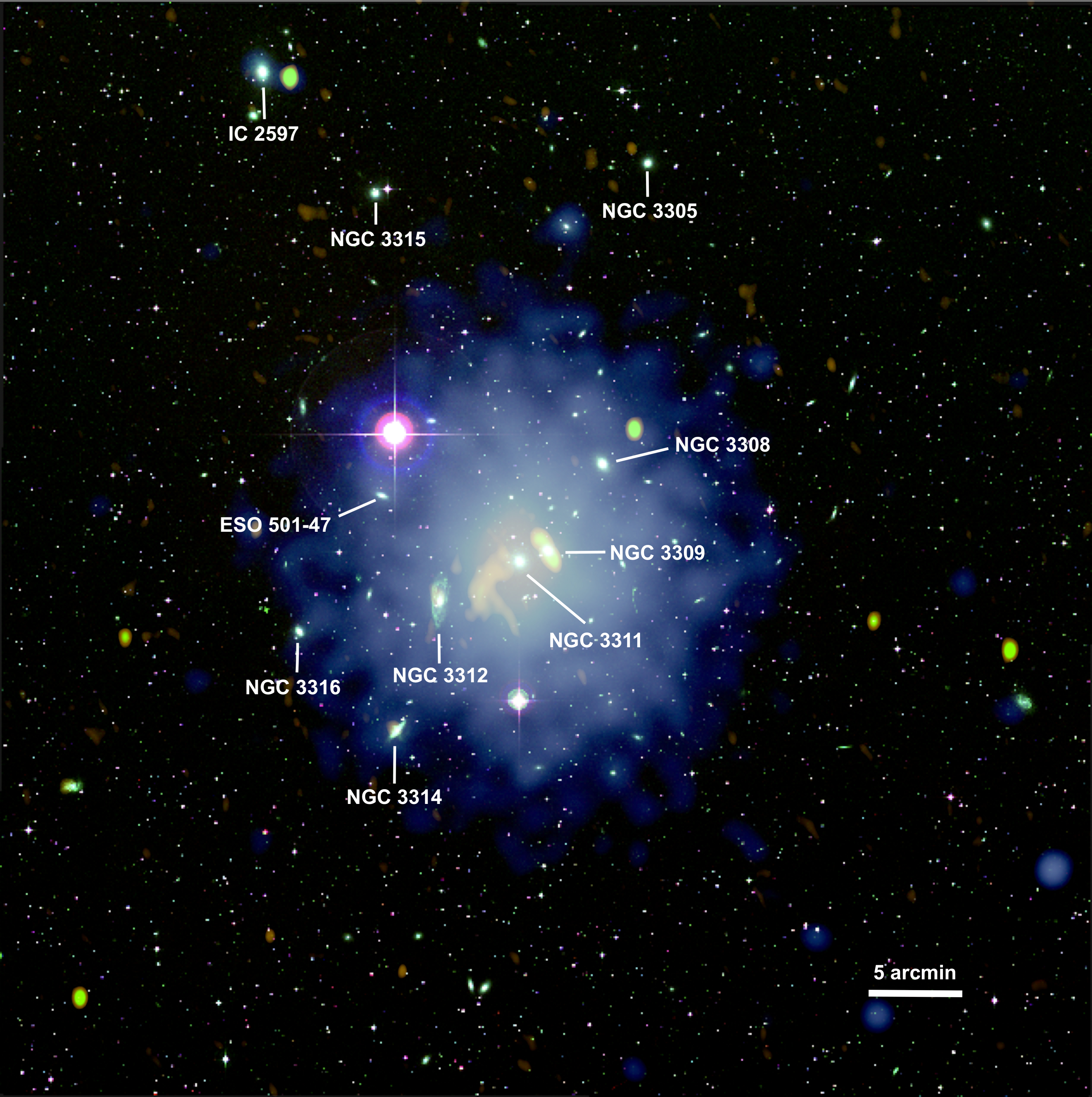}
    \caption{Combined X-ray (cyan-blue), radio (green), and optical/IR (RGB) overlay of the central region ($\approx$0.8\rfive) of Abell 1060. Some of the prominent member galaxies are labeled. The X-ray image is the 0.2--2.3\thinspace\unit{keV} fully corrected TM0 eROSITA image, the radio image is the 150 MHz TGSS image, and the optical image is the DSS2 RGB image (using infrared, red, and blue filters). The $Asinh$ scaling was applied to the X-ray and radio images to enhance them. }
    \label{fig:overlay}
\end{figure}

\subsection{eROSITA spectral analysis}
\label{sec:spectralanalysis}
We performed an in-depth spectral analysis of Abell 1060 to extract the radial distribution of the physical properties of its ICM, for example, gas density, gas temperature, and metal abundance until \rtwo. We used the X-ray spectral fitting software \textsc{XSPEC} \citep{xspec} version 12.12.0 and employed \mbox{C-statistics} \citep{cash} for the fit. Our fitting strategy is based on the eROSITA spectral analyses performed in \mbox{\citet{specghia}}, \mbox{\citet{specliu}}, and \citet{2024veronica,angiecent}. We used the solar abundance tables from \citet{asplund} and the FWC data of the c020 processing version \citep[][Appendix A]{yeungfwc}. \\
\indent
For the analysis, we extracted the source and CXB spectra from within \rtwo\;and eight background regions (\cref{fig:cxbsetup}), respectively, from the cleaned TM0 event list using the eSASS task \texttt{srctool}. The output of this task is the spectrum and response files corresponding to each TM from the extraction region. We then rebinned the spectra to have at least one count per bin using the \texttt{FTOOLS} task \texttt{grppha}. Furthermore, we modeled the combined emission from the ICM, CXB, and PIB using the following model,
\begin{multline}
    \label{eq:clustermodel}
    \texttt{Model}=\underbrace{\texttt{apec}_{\texttt{LHB}} + \texttt{TBabs}* (\texttt{apec}_{\texttt{MWH}} + \texttt{nei}+
    \texttt{powerlaw})}_{\text{CXB}} \\+ \underbrace{\texttt{TBabs} * \texttt{apec}_{\texttt{ICM}}}_{\text{ICM}} + \;\texttt{PIB}\;.
\end{multline}
This spectral model consists of three separate components for the CXB, the ICM, and the PIB and is a modified version of the spectral model from \citet{2024veronica,angiecent}. Our CXB model consists of thermal \texttt{apec} \citep{apec} components for the 
local hot bubble (LHB) and the Milky Way halo (MWH), \texttt{apec}$_\mathrm{LHB}$ and \texttt{apec}$_\mathrm{MWH}$, a constant temperature non-equilibrium ionization model, \texttt{nei},\footnote{\href{https://heasarc.gsfc.nasa.gov/xanadu/xspec/manual/node200.html}{https://heasarc.gsfc.nasa.gov/xanadu/xspec/manual/node200.html}} which models the emission from the Antlia SNR \citep{2026A&A...708A.198K}, and a \texttt{powerlaw}\footnote{\href{https://heasarc.gsfc.nasa.gov/docs/software/xspec/manual/node221.html}{https://heasarc.gsfc.nasa.gov/docs/software/xspec/manual/node221.html}} component to model the emission from the unresolved active galactic nuclei (AGN). We note that the MWH, SNR, and the unresolved AGN components are affected by the Galactic photoelectric absorption along the line of sight and therefore, a \texttt{TBabs} \citep{2000wilms} component is multiplied to them. The $N_\mathrm{H,tot}$ values used for all the \texttt{TBabs} components are extracted from the $N_\mathrm{H,tot}$ map in \cref{fig:nhtotmap}. Subsequently, we used our CXB model and performed a detailed spectral analysis of the CXB in our FoV, which we describe in Appendix \ref{sec:cxbanalysis}. Finally, we modeled the ICM emission also as an absorbed thermal component, \texttt{apec}$_\mathrm{ICM}$, and the \texttt{PIB} component as a set of power laws, an exponential cut-off, and 23 Gaussian fluorescence lines \mbox{\citep[more details in][]{angiecent}}. \\
\indent
    We now describe our fitting procedure to estimate the \mbox{best-fit} normalization, temperature, and metallicity of the \texttt{apec}$_\mathrm{ICM}$ component. This procedure is similar to the one described in \citet{2024veronica,angiecent} with some minor differences. In our CXB analysis (Appendix \ref{sec:cxbanalysis}), we examined the variation in the \texttt{apec}$_\mathrm{LHB}$, \texttt{apec}$_\mathrm{MWH}$, and \texttt{nei} normalizations between all the background boxes and determined that the box in the northwest (\cref{fig:nhtotmap}) is representative of the background within \rtwo. Therefore, we fixed the parameters of the CXB components to the values in \cref{tab:cxb parameters} and freed their normalizations. We then fitted the CXB spectra from each TM simultaneously with our CXB model by keeping the normalizations linked between them. The energy ranges used for TM8 and TM9 are \mbox{0.3--9.0\thinspace\unit{keV}} and \mbox{0.5--9.0\thinspace\unit{keV}}, respectively. Subsequently, we fitted the cluster spectra with the full spectral model (Eq.\thinspace\ref{eq:clustermodel}) by setting the obtained best-fit CXB normalizations as the starting point. This approach accounts for bin to bin variations in the background and propagates them into the cluster fit. The \texttt{nei} normalization was kept unlinked between the cluster and the CXB fitting, as it varies spatially within \rtwo\;and showed some degeneracy with the \texttt{apec}$_\mathrm{LHB}$ in our CXB analysis \mbox{(Appendix \ref{sec:cxbanalysis})}. Finally, the metallicity of the \texttt{apec}$_\mathrm{ICM}$ component was freed and its redshift parameter was set to $z=0.0121$ based on our galaxy distribution analysis (\mbox{Sect.\thinspace\ref{sec:galdistrib}}). To avoid systematic uncertainties associated with the Fe-L shell complex in the low surface brightness regime, where it becomes degenerate with background components and biases metallicity measurements, we omitted the 0.8--1.2 keV band from the fitting \citep[more details in][]{2021A&A...646A..92G}. Since this omission removed a substantial amount of signal, we fixed the normalizations of the \texttt{PIB} component to the best-fit values obtained during the CXB fitting to reduce the model degrees of freedom (d.o.f.). This is validated by \mbox{\citet{angiecent}}, where a minimal variation of $\approx$0.6$\sigma$ is reported in statistical errors when the \texttt{PIB} normalization was frozen. Our best-fit values obtained for ICM normalization, temperature, and metallicity are described in \cref{tab:tempprof}. 

\section{Results and discussion}
\label{sec:results}
\subsection{X-ray images}
\label{sec:xrayimages}
\begin{figure}[h]
    \centering
    \includegraphics[width=0.95\linewidth]{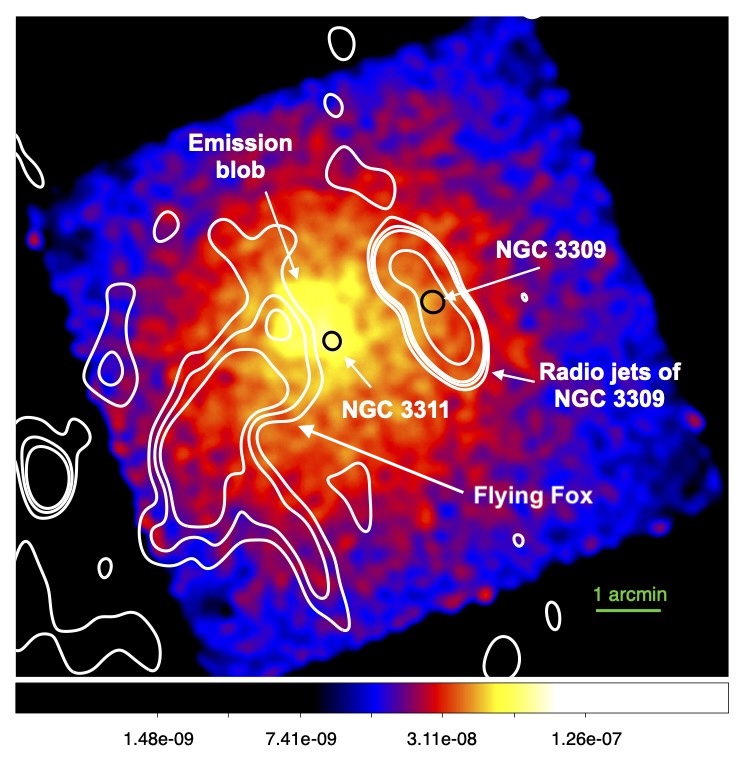}
    \caption{Fully corrected and point source removed \textit{Chandra} image in the 0.5--2.3\thinspace\unit{keV} band overlaid with TGSS radio contours. The image is displayed using a logarithmic scale and is smoothed by a Gaussian kernel of $\sigma=6$ pixels. The black circles mark the excised halos of NGC~3311 and NGC~3309.}
    \label{fig:chandra_TGSS}
\end{figure}
\begin{figure}
    \centering
    \includegraphics[width=1.0\linewidth]{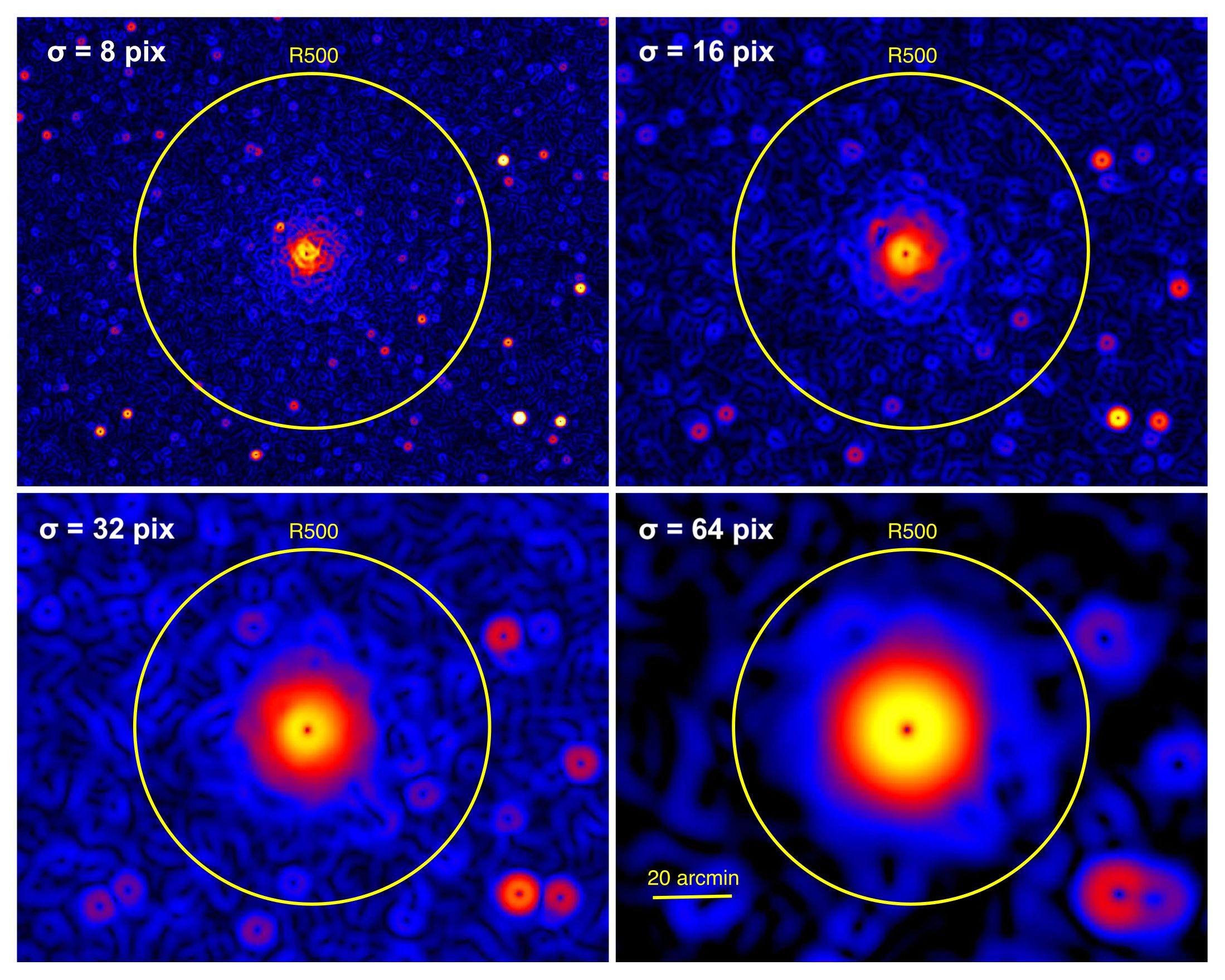}
    \caption{GGM filtered eROSITA images in the 0.2--2.3\thinspace\unit{keV} band. The kernel size is denoted in the top left corner of each image. }
    \label{fig:eroggm}
\end{figure}
\begin{figure}
    \centering
    \includegraphics[width=1.0\linewidth]{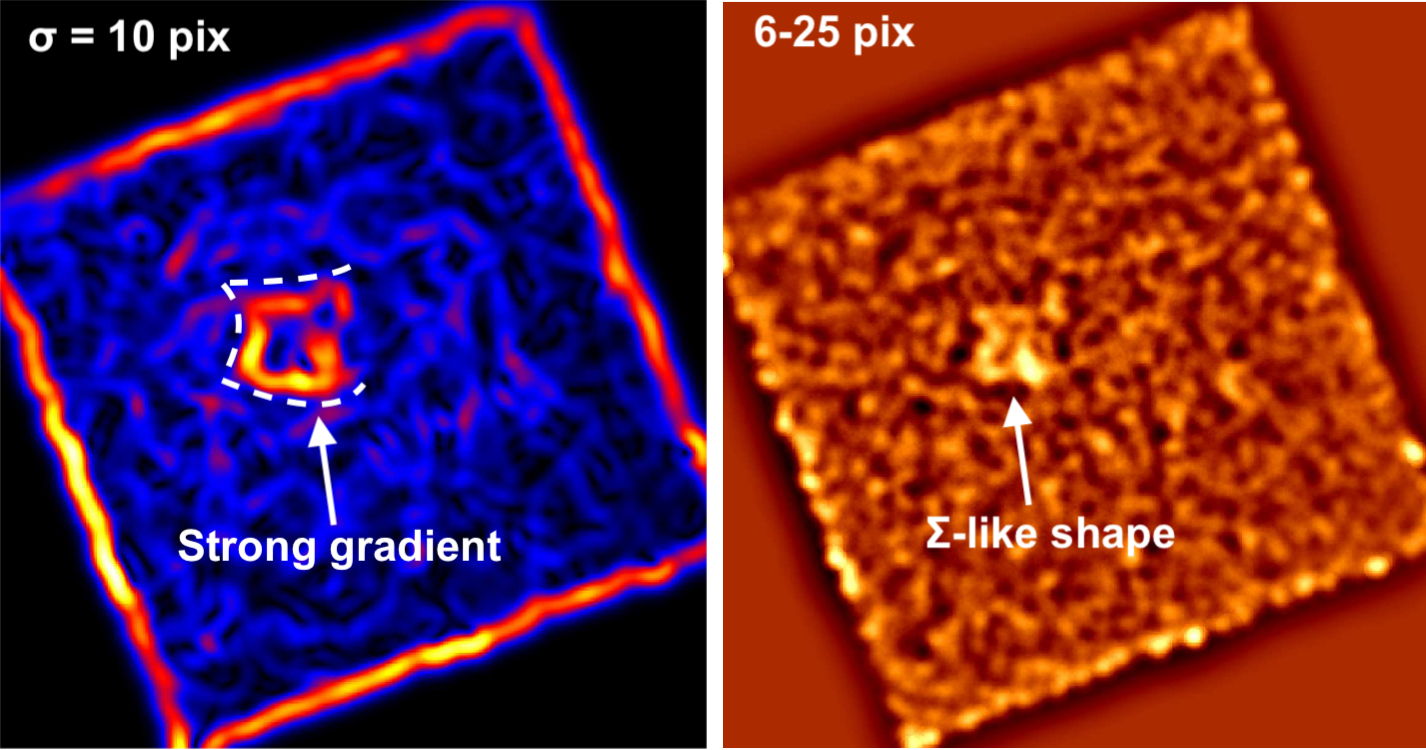}
    \caption{GGM filtered (left) and unsharp masked (right) \textit{Chandra} images in the 0.5--2.3\thinspace\unit{keV} band. The kernel sizes and combinations are denoted in the top left corner of each image.}
    \label{fig:chandraggm+unsharp}
\end{figure}

We present a multiwavelength view of Abell 1060 within $\approx$0.8\rfive\;using the X-ray+radio+optical/IR overlay image in \cref{fig:overlay}. We spatially compared the thermal component of the emission, displayed using the 0.2--2.3~keV eROSITA image and the optical/IR emission, displayed using the \mbox{Digitized Sky Survey 2} \citep[DSS2;][]{2000ASPC..216..145M} infrared, red, and blue images, to establish that the X-ray peak is centered on NGC 3311. The other central galaxy, NGC 3309, is located $\approx\ang[angle-symbol-over-decimal]{;1.5}$ to its west. We also observe the X-ray emission from other prominent members of the cluster, such as the spiral galaxy NGC~3312, NGC~3305, NGC~3308, NGC~3314, NGC~3315, NGC~3316, ESO 501-47, and IC 2597. The overall shape of the X-ray emission from the ICM appears spherically symmetric and shows an undisturbed morphology with a lack of substructures, which is consistent with the previous studies of the cluster (Sect.\thinspace\ref{sec:intro}). We also display the diffuse radio emission from the cluster using the GMRT 150 MHz All-sky \mbox{Radio Survey,} which is a part of the TIFR GMRT \mbox{Sky Survey} \citep[TGSS;][]{tgss} project in \cref{fig:overlay}. We note the presence of the diffuse radio source near NGC 3311, which was discovered by \citet{2024radiohydra} and named the Flying Fox (\cref{fig:chandra_TGSS}). Other bright radio sources are the radio jets of NGC~3309 and the faint radio emission from NGC~3312. We note that the maximum resolvable scale by the TGSS observations is \ang{;68}, which is comparable to Abell~1060's \rtwo\;\citep{tgss}. However, we do not detect any cluster-scale diffuse synchrotron emission, for example, radio halos and relics. In the GGM filtered 0.2--2.3\thinspace\unit{keV} eROSITA images (\cref{fig:eroggm}), we only observe small-scale fluctuations in the $\sigma=8$ and 16 pixels images, and any large-scale shocks or sloshing features with extents of the order of tens or hundreds of \mbox{kiloparsec} are absent. This suggests that unless the last merger happened along the line of sight, it is likely that Abell~1060 has not undergone any major particle reacceleration event in its recent history, and the majority of the diffuse radio emission is indeed due to galaxy-galaxy and galaxy-ICM interactions within $R < \ang{;10}$.  \\
\indent
We investigated this intricate interplay
between the thermal and nonthermal emission within the inner \ang{;8}$\times$\ang{;8} region of the cluster using the point source removed \mbox{0.5--2.3\thinspace\unit{keV}} \textit{Chandra} image with the TGSS radio contours overlaid (\cref{fig:chandra_TGSS}). We identify that the stripped halo of NGC~3311 (Sect.\thinspace\ref{sec:intro}), labeled as \lq\lq Emission blob\rq\rq\;in the image, has three sharp surface brightness edges toward the south, east, and north. The eastern edge overlaps with the northernmost part of the Flying Fox in projection. We further validated these features using the GGM filtered \mbox{0.5--2.3\thinspace\unit{keV}} \textit{Chandra} image (\cref{fig:chandraggm+unsharp} left), where strong gradients along the southern, eastern, and northern surface brightness edges are visible. These gradients together exhibit a \lq\lq$\Sigma$\rq\rq-shaped structure in the unsharp masked \textit{Chandra} image (\cref{fig:chandraggm+unsharp} right). We performed a surface brightness analysis to quantitatively analyze these edges, which is described in Sect.\thinspace\ref{sec:sbshock}. In addition, contrary to NGC 3311, NGC~3309 shows no signs of extended X-ray emission beyond its excised halo. Furthermore, we do not detect any cavities near its radio jets, which suggests a weak interaction between the jets and the ICM, and its AGN is likely in a quiescent phase. However, this conclusion is limited by the \mbox{signal-to-noise} ratio (S/N) of the current observation \citep[e.g.,][]{2010ApJ...712..883D,2016ApJS..227...31S}. 

\begin{figure}[h] 
    \centering
    \includegraphics[width=0.99\linewidth]{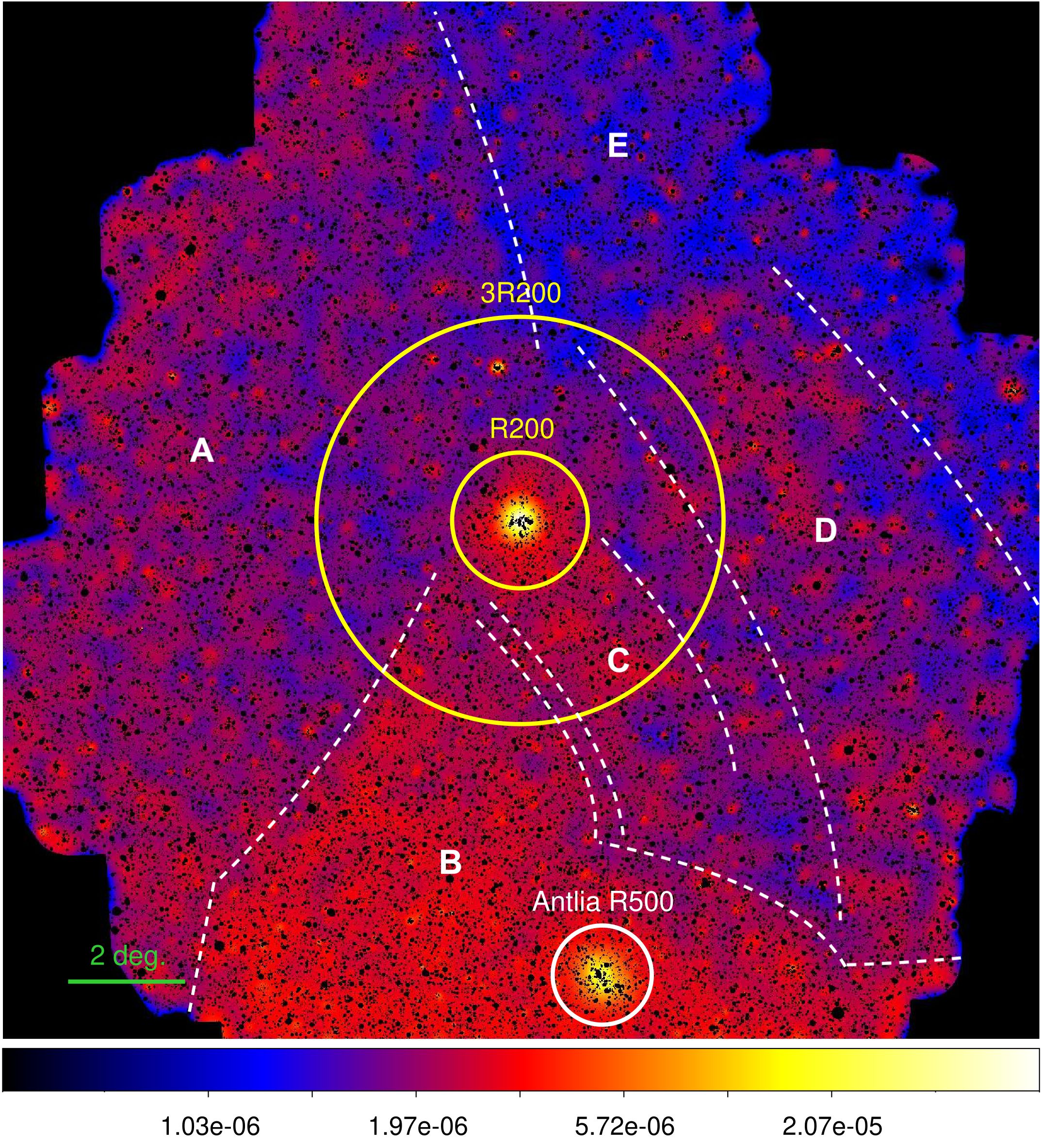}
    \caption{Point source removed eRASS:4 TM0 wavelet filtered image in the \mbox{0.2--2.3\thinspace\unit{keV}} band. The image is plotted using a logarithmic scale and is in units of counts\thinspace\unit{s^{-1}}. The characteristic radii of Abell 1060 (from \cref{tab:clusterparams}) and Antlia \citep[from][]{2016antlia} are overlaid. The distinct foreground structures in the FoV are labeled from A to E.}
    \label{fig:fc_wf}
\end{figure}
We also present the first large-scale view of the X-ray emission from Abell 1060, the Antlia cluster, and the foreground structures in our FoV using the \mbox{0.2--2.3\thinspace\unit{keV}} eROSITA wavelet filtered image (\cref{fig:fc_wf}). We observe a considerable amount of foreground emission beyond Abell 1060's \rtwo\;that extends predominantly toward the south and west. This emission has high spatial variability, and its various substructures are visibly distinguishable. Therefore, we visually separated these structures, including the emission from the Antlia SNR, into five distinct regions and labeled them from A to E (\cref{fig:fc_wf}). Out of these regions, region B contains the majority of the emission from the Antlia SNR \citep{2026A&A...708A.198K}, which obscures the Antlia cluster. Therefore, it is extremely difficult to visually distinguish any extended filamentary emission between the two clusters that might be in the background. We further observe that regions A and D show noticeably lower emission as compared to region B, but are highly extended beyond Abell~1060's 3\rtwo. On the contrary, the majority of the emission in region C is localized between \rtwo\;and 3\rtwo. Finally, region E, which is located at the northernmost edge of our FoV, shows a visible lack of foreground structures. From \cref{fig:RGBhalfsky}, we further note that the \ero\;bubble is >\ang{10} to the east of our FoV, and thus too distant to cause any significant effect on the X-ray foreground. We also notice that the foreground structures in regions A, B, and D extend farther beyond our FoV to the south and west, and are prominent in the 0.4--1.0\thinspace\unit{keV} range due to their \mbox{reddish-green} hue. Hence, we can distinguish between the various foreground structures as a function of energy using our eROSITA RGB image in \cref{fig:wfrgb}. We note that the emission in region B has predominantly a soft spectrum because of its \mbox{reddish-green} color, with the \mbox{0.8--1.2\thinspace\unit{keV}} component increasing as we move toward the Antlia cluster. Other foreground features throughout the FoV also exhibit a similar composition with varying contributions of 0.2--0.8\thinspace\unit{keV} and 0.8--1.2\thinspace\unit{keV} components. The harder \mbox{1.2--2.3\thinspace\unit{keV}} component is limited only to the emission from the two clusters and other extended background sources.  \\
\begin{figure}[h] 
    \centering
    \includegraphics[width=0.99\linewidth]{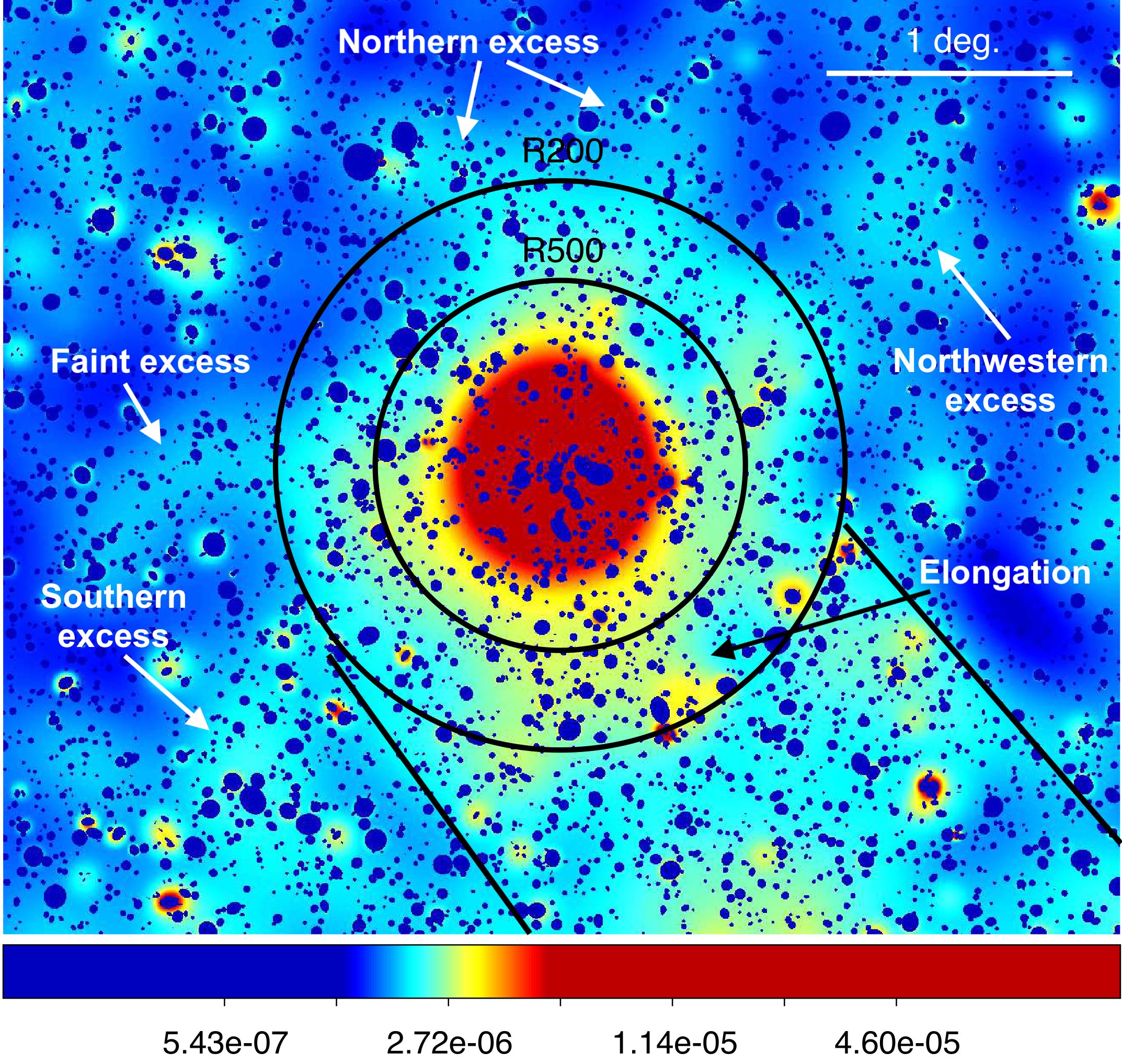}
    \caption{Same as \cref{fig:fc_wf} but displayed using a different color map and zoomed in on Abell 1060. The prominent X-ray features are labeled, including the foreground emission (between solid black lines) from region C.}
    \label{fig:wfclosup}
\end{figure}
\indent
In \cref{fig:wfclosup}, we show the enhanced soft X-ray emission that extends beyond the \rtwo\;of Abell 1060. Toward the north, we see a bidirectional, low surface brightness excess that further bifurcates toward the northeast and northwest. Similar excesses are also present in the northwest and east. Additionally, we observe a  brighter southern excess that is located toward the northernmost corner of region B. However, we suspect that this excess is highly contaminated by the foreground emission from the Antlia SNR. Furthermore, we notice a surface brightness excess between \rfive\;and \rtwo\;in the south that is possibly in part connected to the foreground structure in region C (\cref{fig:wfclosup}). We further validate the presence of these excesses via our sector surface brightness analysis in Sect.\thinspace\ref{sec:sectorsb}.
\indent

\subsection{Galaxy distribution}
\label{sec:galdistrib}
\begin{figure}
    \centering
    \includegraphics[width=0.99\linewidth]{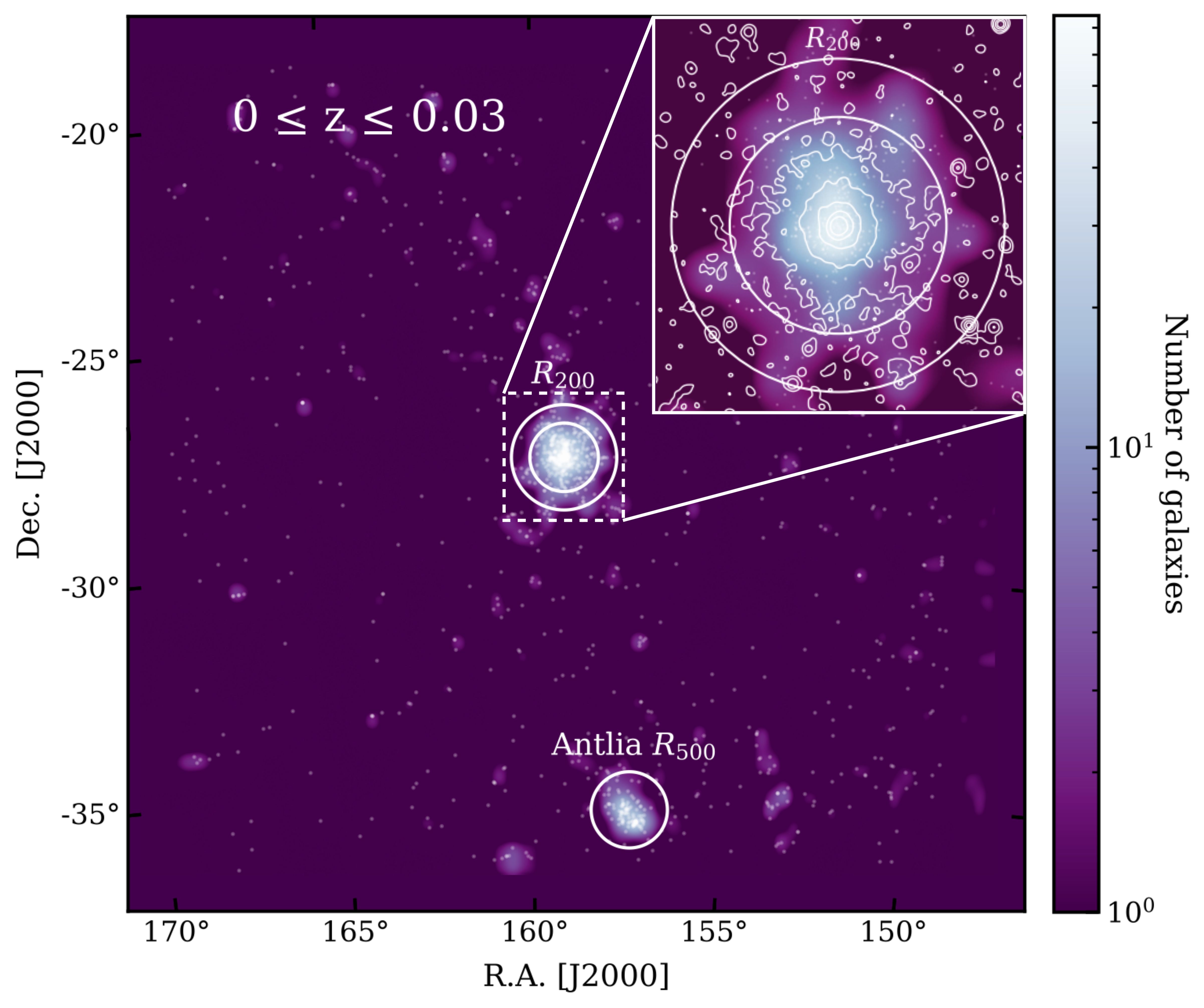}
    \caption{Galaxy distribution from the NED galaxy catalog reprojected to match our eROSITA FoV, within the redshift range $0\leq z \leq 0.03$. The inset in the top right corner is the same image zoomed in on Abell~1060's \rtwo\;and overlaid with the eROSITA 0.2--2.3\thinspace\unit{keV} X-ray contours. The circle inside \rtwo\;represents the \rfive.}
    \label{fig:NED_Xray}
\end{figure}

We performed an optical galaxy redshift analysis of Abell~1060 using the NASA/IPAC Extragalactic Database (NED)\footnote{\href{https://ned.ipac.caltech.edu/byparams}{https://ned.ipac.caltech.edu/byparams}} galaxy catalog with a redshift upper limit of $z=0.1$ and an extent that matches our eROSITA FoV. The redshifts in this catalog are corrected to the Cosmic Microwave Background rest frame, and the breakdown of redshifts in this catalog between spectroscopic, photometric, and redshifts from unknown sources is 72.74\%, 6.66\%, and 20.44\%, respectively. 
\begin{figure}
    \centering
    \includegraphics[width=1.0\linewidth]{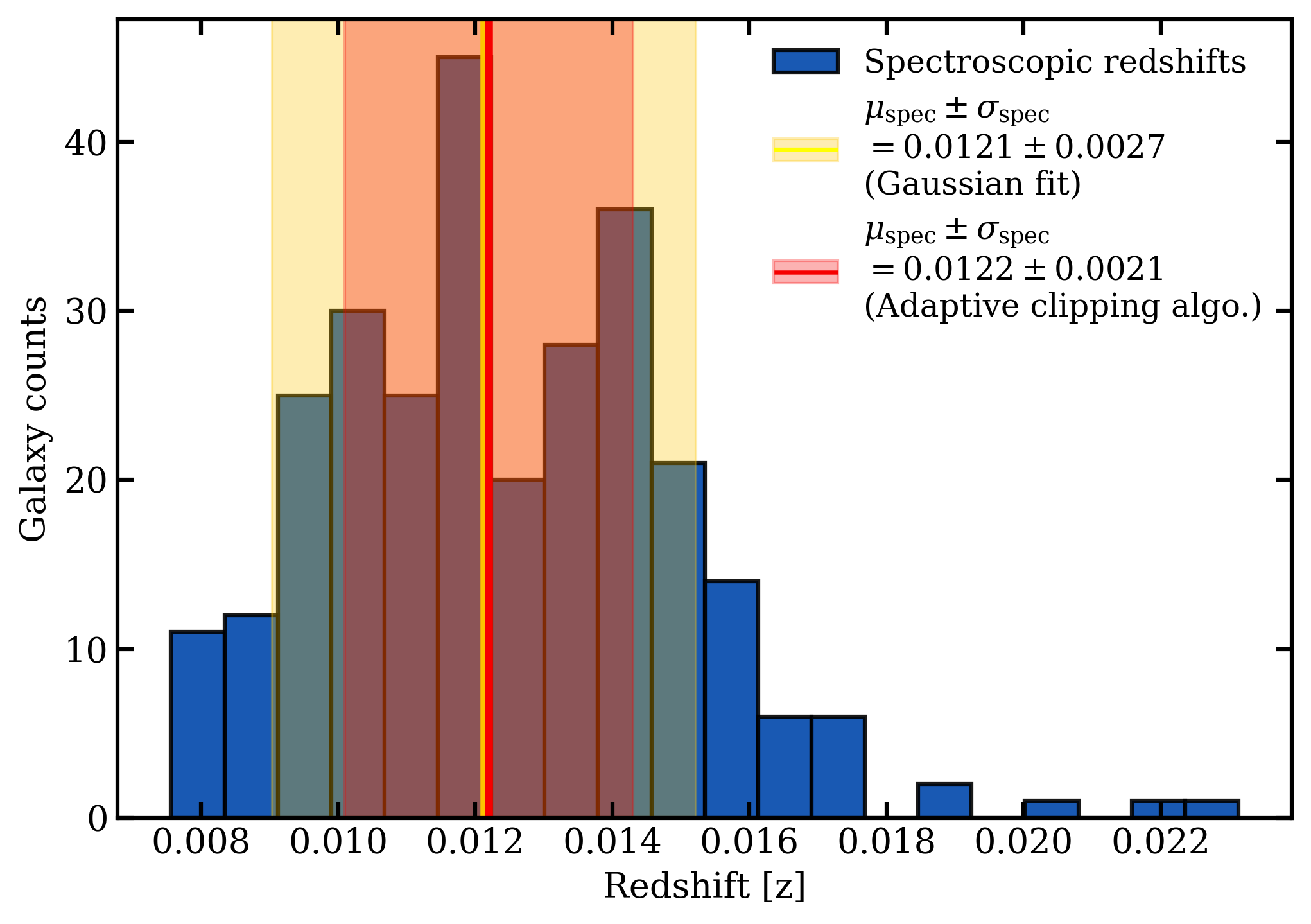}
    \caption{NED spectroscopic redshift distribution within \rtwo. The obtained mean and standard deviation of the distribution are represented by the solid lines and shaded regions, respectively.}
    \label{fig:galdistr}
\end{figure}
For the cluster redshift estimation, we only included galaxies with spectroscopic redshifts within \rtwo\;and $\approx$$5\sigma$ range from the MCXC redshift of $z=0.0126$ (corresponding redshift range $0\lesssim  z\lesssim 0.03$), given Abell 1060's line of sight galaxy velocity dispersion of $\sigma_\mathrm{v}\approx700\;\mathrm{km\;s}^{-1}$ \citep[e.g.,][]{2011A&A...531A...4M}. In \cref{fig:galdistr}, we fitted a Gaussian function to this unimodal spectroscopic redshift distribution and obtained a best-fit mean ($\mu_{\mathrm{spec}}$) and standard deviation ($\sigma_{\mathrm{spec}}$) of 0.0121 and 0.0027, respectively. We also applied an adaptive $n\sigma$-clipping algorithm to this distribution, which yielded a nearly identical redshift estimate of \mbox{$\mu_{\mathrm{spec}} \pm \sigma_{\mathrm{spec}} =0.0122\pm0.0021$.} In both cases, the obtained $\sigma_\mathrm{spec}$ is broadly consistent with the literature velocity dispersion of Abell~1060. Moreover, the apparent unimodality in Abell~1060's redshift distribution is in line with the lack of substructure observed in the 0.2--2.3\thinspace\unit{keV} eROSITA image (\cref{fig:wfclosup}) within \rfive. We further investigated the correlation between the X-ray emission and the 2D galaxy distribution in our FoV using the 2MASS galaxy density contours \citep{2mass,2massreip} and the NED catalog. In \cref{fig:NED_Xray}, the galaxy distribution within \rfive\;shows high spatial correlation with the centrally peaked \mbox{X-ray} emission from the ICM, which is typical for a relaxed cluster. Furthermore, we identified two arm-like galaxy overdensities in \cref{fig:2MASS_wf} that extend beyond \rtwo\;in the west and southeast. The southeastern overdensity further extends by $\approx$\ang{5} to the south beyond 3\rtwo\;and overlaps with the foreground structure in region B. However, both these structures consist mainly of background eRASS:1 clusters \citep[e.g.,][]{2024A&A...685A.106B} and galaxies in the redshift range \mbox{$0.05 \leq z \leq 0.08$}, which are unrelated to Abell~1060's galaxy distribution. Nevertheless, several \mbox{small-scale} overdensities closer to \rtwo\;in the north, \mbox{northwest}, \mbox{southwest}, and southeast are visible in \cref{fig:NED_Xray}, which implies that at least a fraction of the emission from the southern extended excesses in \cref{fig:wfclosup} belongs to the cluster.  \\ 
\indent
Furthermore, we find no evidence of a galaxy overdensity directly connecting Abell 1060 and the Antlia cluster within
the redshift range \mbox{$0 \leq z \leq 0.03$}. However, several indirect overdensities that extend beyond Abell 1060's \rtwo\; and eventually connect to Antlia's galaxy distribution are apparent in \cref{fig:2MASS_wf}, which are in the redshift range $0.03\leq z \leq 0.08$ in the NED catalog. Despite this, the validation of these features using our X-ray images is challenging since the relatively high emission measure (EM) foreground emission from the Antlia SNR has a similar temperature \citep[0.15--0.18\thinspace\unit{keV};][]{2026A&A...708A.198K} as compared to the possible filamentary X-ray emission from the warm-hot intergalactic medium \citep[WHIM, $\sim$0.01--0.9\thinspace\unit{keV}; e.g.,][]{1999ApJ...514....1C,2024jakob} that we intend to detect. 

\subsection{X-ray surface brightness analysis}
\label{sec:sbanalysis}
\begin{figure}[t]    
    \includegraphics[width=1.0\linewidth]{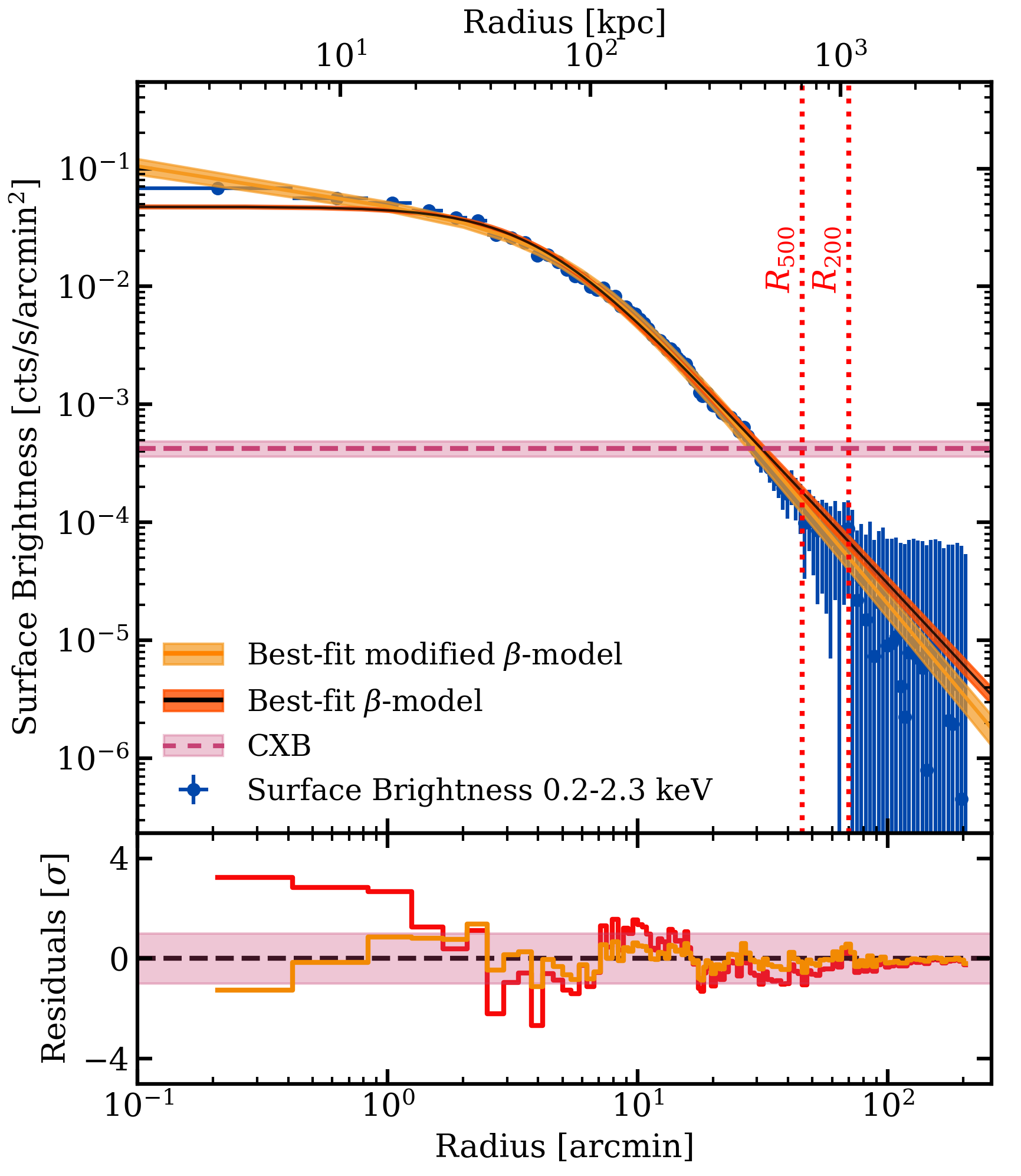}
    \caption{CXB subtracted eROSITA surface brightness profile of Abell~1060 in the 0.2--2.3\thinspace\unit{keV} band. Also plotted are the best-fit models (orange and red) and average CXB level (pink) with their respective $1\sigma$ uncertainties (shaded regions). In the residual plot, the region where the residuals are less than $1\sigma$ is shaded in pink.}
    \label{fig:SBprofile}
\end{figure}

We extracted the eROSITA surface brightness profile of Abell~1060 until 3\rtwo\;in the 0.2--2.3\thinspace\unit{keV} band, following the procedure described in \citet[technical details are provided in our \cref{app:sbanalysis}]{angiecent}. The profile is centered on the X-ray peak, \mbox{R.A. = \ang[angle-symbol-over-decimal]{159.184}} and Dec.~=~\ang[angle-symbol-over-decimal]{-27.527}, which is estimated within an aperture of $R=\ang{;10}$ centered on Abell~1060's core in the fully corrected 0.2--2.3\thinspace\unit{keV} eROSITA image. Furthermore, we used eight background boxes of dimensions $\ang{3}\times\ang{1}$, each located at $R=4.5$\rtwo\;(\cref{fig:cxbsetup}) to estimate the average CXB that is representative of the local X-ray background of Abell~1060. We account for both the statistical uncertainty from each box and the dispersion due to the spatial variation of the foreground between all the boxes by taking the mean and standard deviation of the individual surface brightness values as the average CXB and its uncertainty, respectively. Using this approach, we estimated an average CXB level of $(4.25\pm0.63)\times 10^{-4}$\thinspace\unit{counts.s^{-1}.arcmin^{-2}}. In \cref{fig:SBprofile}, we display the CXB subtracted eROSITA profile in the 0.2--2.3\thinspace\unit{keV} band. We also point out that the error bars beyond \rtwo\;are dramatically enlarged due to poor statistics and the surface brightness being nearly identical to the CXB level.

To model the radial surface brightness distribution of the ICM, we fitted the profile with a modified version of the single \mbox{$\beta$-model} \citep[e.g.,][]{Cav&FucBeta1976} and the \mbox{AB-model} \citep[e.g.,][]{2002A&A...394..375P} that has the functional form
\begin{equation}
    S_\mathrm{X} (R)=S_\mathrm{X} (0) \left(\frac{R}{r_\mathrm{c}}\right)^{-\alpha}
    \left[ 1+ \left(\frac{R}{r_\mathrm{c}}\right)^2\right]^{-3\beta + \frac{\alpha}{2}+\frac{1}{2}},
    \label{eq:modbetamodel}
\end{equation}
\noindent
where $S_\mathrm{X}(0)$ is the normalization, $R$ is the projected radius, $r_\mathrm{c}$ is the core radius, $\alpha$ is the slope of the central power law component, and $\beta$ is the slope of the overall profile. We used the \texttt{emcee} Markov chain Monte Carlo sampler \citep{emcee} to estimate the posterior distributions of the model parameters, assuming flat priors and a Gaussian likelihood function. The best-fit single and modified $\beta$-models are displayed in \cref{fig:SBprofile} and their model parameters are described in \cref{tab:sbparams}. In the inner three bins, we observe residuals of magnitude $\approx$3.5$\sigma$ from the single $\beta$-model created by the weak cusp in the profile (Sect.\thinspace\ref{sec:intro}). Given the evidence for a relaxed ICM morphology in Sect.~\ref{sec:xrayimages}, the presence of this central cusp is expected. Previously, this excess was modeled using an NFW profile \citep[e.g.,][]{2000hydra} and a double $\beta$-model \citep[e.g.,][]{2002hydra,2002newchandra,2007hydra}. However, the extent of the profiles in these studies was limited to \rfive\;and could not capture the entire surface brightness distribution. In our case, we modeled the full extent of the surface brightness distribution until 3\rtwo\;and determined that both these models failed to fit it. On the other hand, Eq.\thinspace\ref{eq:modbetamodel} proves to be a good fit ($\chi_\mathrm{red}^2 =0.9$) and successfully models the cusp (mean residual of $-0.67$$\sigma$ within $R=\ang{;1}$) with a power law of slope of $\alpha=0.33\pm0.03$. Interestingly, weak surface brightness cusps with $\alpha < 0.5$ are a typical characteristic of NCC clusters \citep[e.g.,][]{2007hvcg.conf...48V}. 

\begin{table}[h]
   \centering
   \caption{Best-fit parameters of the single and modified $\beta$-models for the 0.2--2.3\thinspace\unit{keV} band.}
   \label{tab:sbparams}
   \begin{tabular}{@{}ccc@{}}
   \toprule
    Parameters  & Single $\beta$-model & Modified $\beta$-model \\ \midrule
   $S_\mathrm{X}(0)^{(a)}$ & $(2.71\pm0.20)\times 10^{-2}$&$(4.74\pm0.12)\times 10^{-2}$    \\
   $r_{\text{c}}$ [arcmin]&$3.97\pm0.12$ &$5.91\pm0.33$  \\
   $\alpha$ &-& $0.33\pm0.03$ \\
   $\beta$ &$0.55\pm0.01$&$0.59\pm0.01$  \\
   $\chi_\mathrm{red}^2$&1.58&0.90 \\
   CXB$^{(a)}$&\multicolumn{2}{c}{$(4.25\pm0.63)\times 10^{-4}$} \\
   \bottomrule
   \end{tabular}
   \tablefoot{
      \tablefoottext{a}{Units are counts s$^{-1}$ arcmin$^{-2}$.}
   }
\end{table}
\indent
Furthermore, \citet{2004hydra} previously estimated best-fit values of $\beta=0.56\substack{+0.07 \\ -0.05}$ and $r_\mathrm{c}=\ang[angle-symbol-over-decimal]{;5.5}\substack{+\ang[angle-symbol-over-decimal]{;1.4} \\ -\ang[angle-symbol-over-decimal]{;1.3}}$ using a core excised \textit{Chandra} profile ($\ang{;4}\leq R \leq\ang{;18}$), which agrees with our $\beta$ and $r_\mathrm{c}$ estimates within error bars. \citet{2000hydra} and \citet{2007A&A...466..805C} used ROSAT surface brightness profiles that extended up to 1.1\rfive\;but reported slightly different $\beta$ and $r_\mathrm{c}$ values. \citet{2000hydra} slightly underestimated the parameter values, while \citet{2007A&A...466..805C} estimated \mbox{$\beta=0.61\substack{+0.04 \\ -0.03}$} and $r_\mathrm{c}=\ang[angle-symbol-over-decimal]{;6.08}\substack{+\ang[angle-symbol-over-decimal]{;0.97} \\ -\ang[angle-symbol-over-decimal]{;0.78}}$, which are consistent with our estimates within error bars. Additionally, \citet{2008burns} used N-body and hydrodynamical simulations to obtain a similar $\beta$ value of $\approx$0.66 for both CC and NCC clusters and $r_\mathrm{c}$ values of $(0.05\pm0.09)$\rtwo\;and $(0.12\pm0.02)$\rtwo, respectively. Thus, Abell~1060 again appears to be closer to an NCC cluster on the basis of our $r_\mathrm{c}$ value. We also notice a small kink immediately beyond \rtwo\;in the profile, whose investigation along with the outskirts of Abell 1060 is described in Sect.\thinspace\ref{sec:sectorsb}.

\subsubsection{Sector profiles and outskirts}
\label{sec:sectorsb}
We investigated the directional surface brightness features by splitting the full annulus eROSITA profile into eight sectors, each having an opening angle of $\Delta\varphi=\ang{45}$. We used a similar binning setup as the full annulus profile, but reduced the number of bins between \rfive\;and 3\rtwo\;to 35 to improve the S/N. In \cref{fig:sectorsb}, we present the sector surface brightness profiles and the CXB level for each sector, which are plotted together with the full annulus profile. In these sector profiles, we observe some inhomogeneities in the inner core ($R\leq\ang{;2}$) and only minor statistical fluctuations in some bins between 0.22\rfive\thinspace$\leq R \leq\;$\rfive\;(more details in \cref{app:sbanalysis}). Overall, the projected ICM emission from Abell 1060 is highly uniform on large scales up to \rfive. On the other hand, the outskirts of Abell~1060 are notably complex as most sectors exhibit strong relative variations in surface brightness beyond \rfive. \\
\begin{figure}[H]
    \centering
    \includegraphics[width=1.0\linewidth]{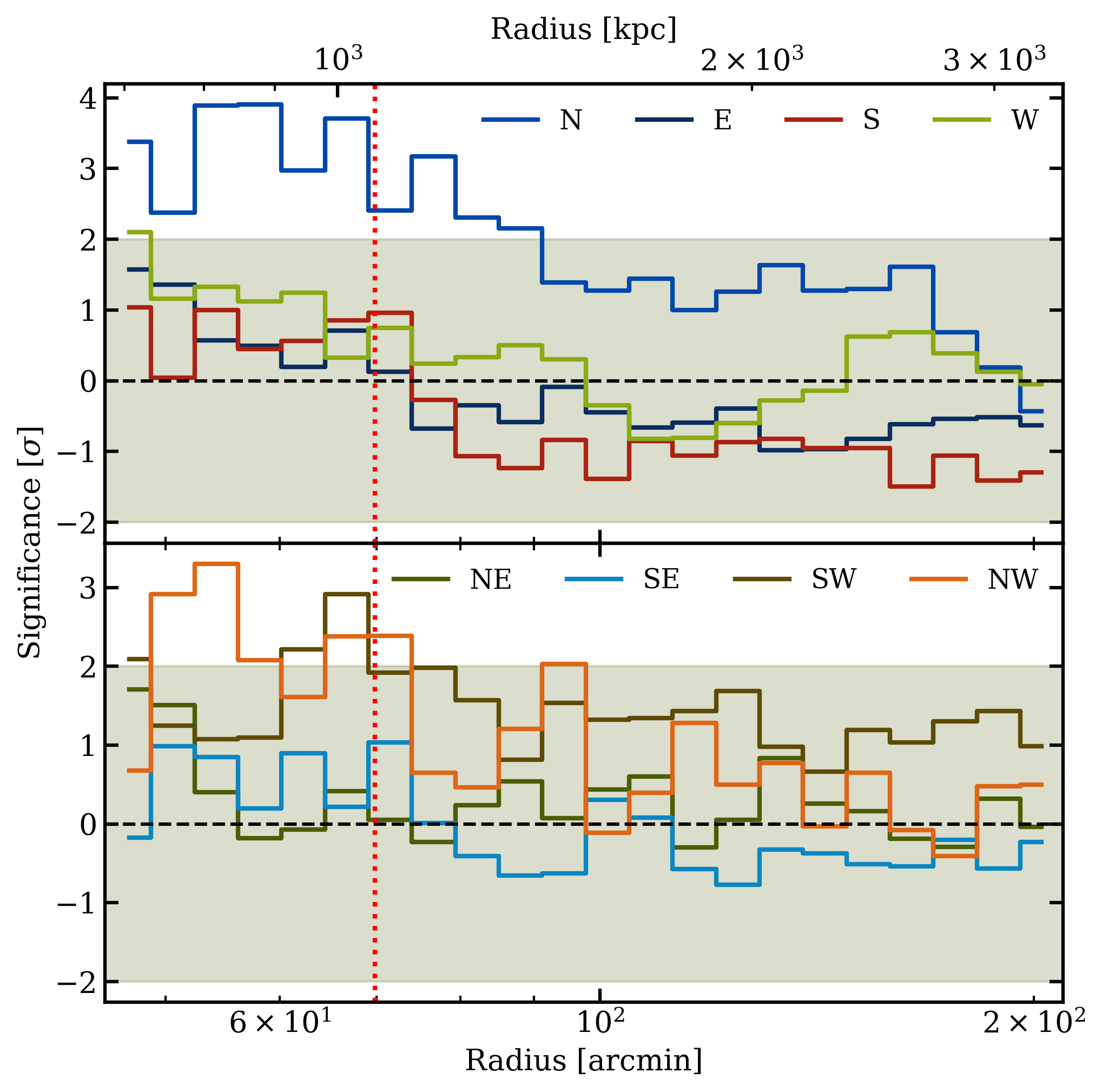}
    \caption{Surface brightness significance profiles of all eight sectors in the 0.2--2.3\thinspace\unit{keV} band. The shaded region represents significance $\in [-2\sigma,2\sigma]$ and the dotted vertical line represents the \rtwo.  } 
    \label{fig:sectorsignif}
\end{figure}
\indent
In \cref{fig:sectorsignif}, we display the surface brightness significance (in units of $\sigma$) above the CXB level, estimated using Eq.\thinspace\ref{eq:sbsignif}, for all sectors beyond \rfive. To account for spatial foreground/background variations, we adopted the nominal CXB value from the background box aligned with each sector. The associated uncertainty was estimated as the standard deviation of the nominal surface brightness values measured in that box and its four neighboring boxes. We treat this setup as the fiducial background estimate, while additionally considering conservative and optimistic cases where the uncertainty is estimated using all background boxes and only the three nearest background boxes, respectively. However, we note that the conservative estimate likely overestimates the true local CXB uncertainty for the three northern sectors, as the Antlia SNR predominantly affects the southern sectors. The kink observed in the full-annulus profile is only present in the southern sector (\cref{fig:sectorsb,fig:sectorsignif}) and arises from the transition between the southern elongation (\cref{fig:wfclosup}) and the foreground emission associated with the Antlia SNR and region C. We also confirm the presence of three soft X-ray excesses in the north, northwest and southwest that extend beyond \rtwo\;(\cref{fig:wfclosup}), with peak fiducial significances of 3.9$\sigma$, 3.3$\sigma$, and 2.9$\sigma$, respectively. Under the conservative and optimistic background treatments, these values vary between 2.9--4.3$\sigma$, 2.4--3.3$\sigma$, and 2.8--2.9$\sigma$, respectively. All of these excesses show a strong spatial correlation with the 2D galaxy distribution (Sect.\thinspace\ref{sec:galdistrib}). Furthermore, the northern and northwestern excesses remain detectable out to 1.4\rtwo\;and 1.1\rtwo\;respectively, with median fiducial significances of 3$\sigma$ and 2.4$\sigma$ (\cref{fig:sectorsignif}). The corresponding conservative and optimistic significance ranges are 2.1--3.4$\sigma$ for the north and 1.7--2.4$\sigma$ for the northwest. This is the farthest any significant ICM emission has been detected from Abell~1060. Similar detections in the outskirts have also been reported for the Centaurus and Fornax clusters using eROSITA \citep[e.g.,][]{angiecent,fornax}. Moreover, this result is consistent with the identification of a virialized northern overdensity using a projected \mbox{phase-space} diagram by \citet{2024A&A...689A.306S} within $R\approx1.5$\rtwo. This further suggests that the north and northwest are preferred directions of active accretion in projection. In comparison, the median fiducial significance in the southwest between \rfive\;\thinspace and 3\rtwo\;\thinspace is only 2$\sigma$, varying between 1.9$\sigma$ and 2$\sigma$ under the conservative and optimistic background treatments, respectively. A clear distinction between the ICM and the extended foreground structure in region C is therefore difficult to make despite the presence of a galaxy overdensity beyond \rtwo. Nevertheless, the robust excesses detected in the north and northwest, together with the tentative southwest excess, suggest that Abell 1060 is actively accreting baryons along multiple directions and that its outskirts are currently being assembled. Meanwhile, all other sectors remain largely within 2$\sigma$ of the CXB level until 3\rtwo. \citet{2015ApJ...810...36M} showed that the inner regions of a cluster usually virialize before the outskirts cease to accrete, which is consistent with our identification of multiple X-ray excesses beyond \rtwo, a relaxed ICM (Sect.\thinspace\ref{sec:xrayimages}), and the weak central cusp (Sect.\thinspace\ref{sec:sbanalysis}) in Abell~1060. 

\subsubsection{Surface brightness discontinuities}
\label{sec:sbshock}
We present the surface brightness analysis of the edges of the ram-pressure stripped halo of NGC 3311 (\cref{fig:chandraggm+unsharp}), which we performed using the \texttt{pyproffit} \citep{pyproffit} package and the 0.5--2.3\thinspace\unit{keV} \textit{Chandra} image. We extracted the surface brightness profiles within $R=\ang{;2}$ from the eastern, southern, and northern directions with an opening angle of $\Delta\varphi = \ang{70}$, based on the visual prominence of the gradients in these directions in \cref{fig:chandraggm+unsharp}. Furthermore, we iteratively adjusted the position and binning of the annuli to ensure that their center is at the X-ray peak orthogonal to the surface brightness discontinuity, and the feature is properly sampled in the profile. We then modeled the surface brightness profiles to characterize the density contrasts across the discontinuities using a broken power law model \citep[e.g.,][]{2023ApJ...944..132S} of the form
\begin{equation}
\label{eq:sbjump}
    S_\mathrm{X}(R)=S_\mathrm{X}(0)\int F(\omega)^2\;\mathrm{d}l,
\end{equation}
where $\omega^2=R^2+l^2$, and
\begin{equation*}
\label{eq:densityprofile}
    F(\omega)=
    \begin{cases}
      \;\omega^{-\alpha_1}, & \omega<r_{\text{f}}\\
      \;\frac{1}{J}\omega^{-\alpha_2}, & \omega\geq r_{\text{f}}
   \end{cases} 
   \;,
\end{equation*}
where $F(\omega)$ is the density profile, $\omega$ is the 3D radius, $l$ is the distance along the line of sight, $\alpha_1$ and $\alpha_2$ are the power law indices, $J$ is the density contrast, i.e.,~$\rho_2/\rho_1$, and $r_{\text{f}}$ is the position of the discontinuity.
\begin{figure}[h]
    \centering
    \includegraphics[width=1.0\linewidth]{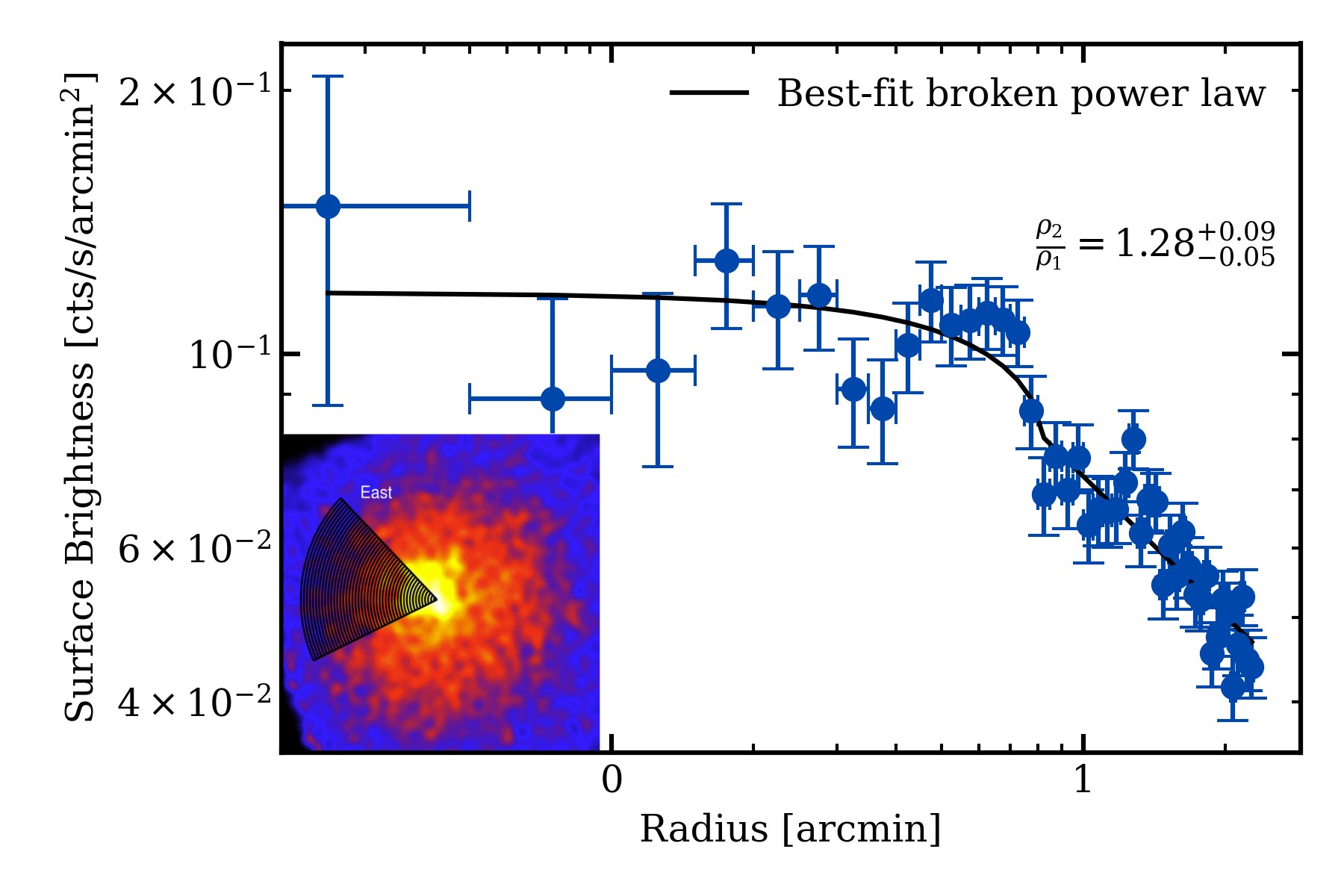}
    \includegraphics[width=1.0\linewidth]{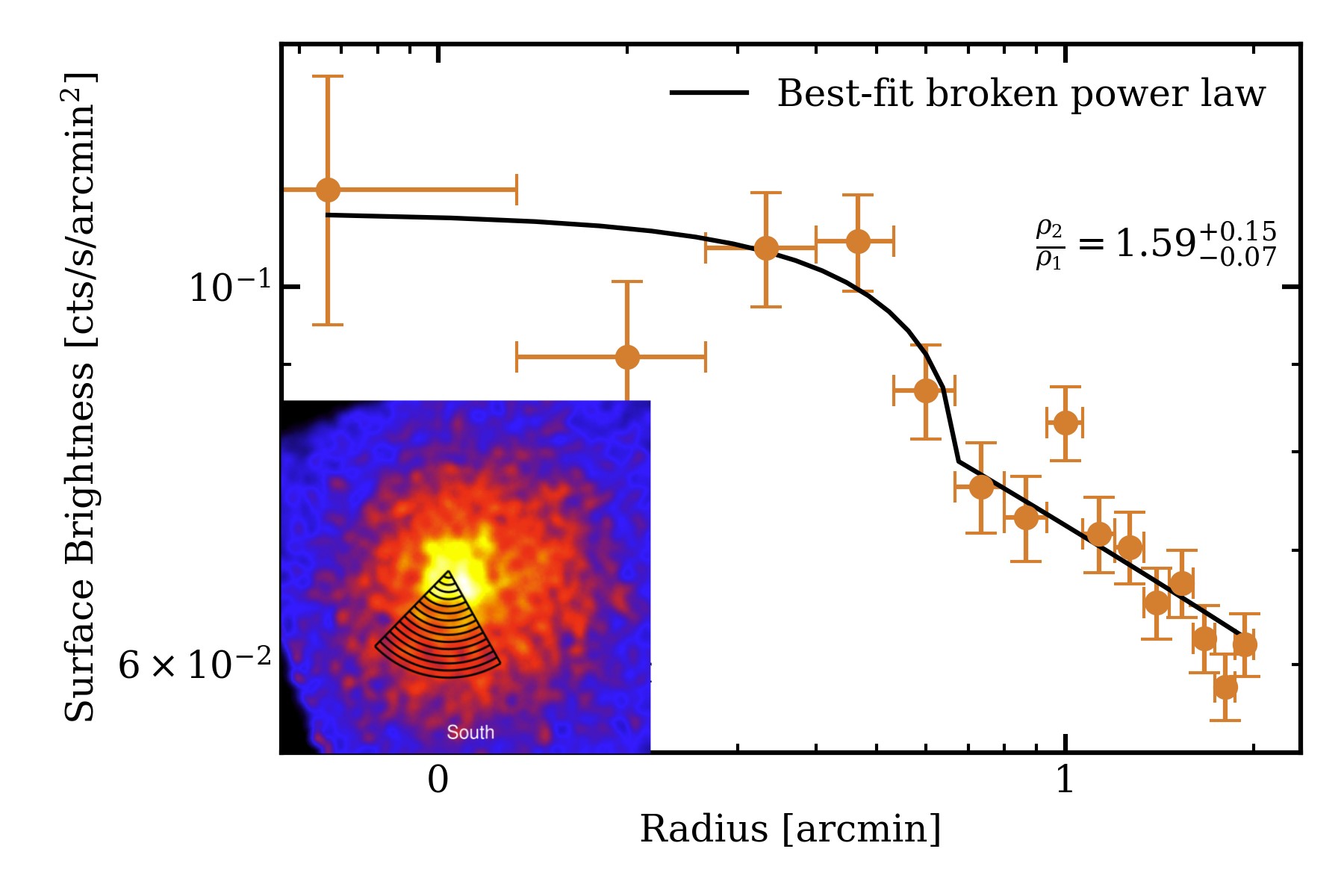}
    \includegraphics[width=1.0\linewidth]{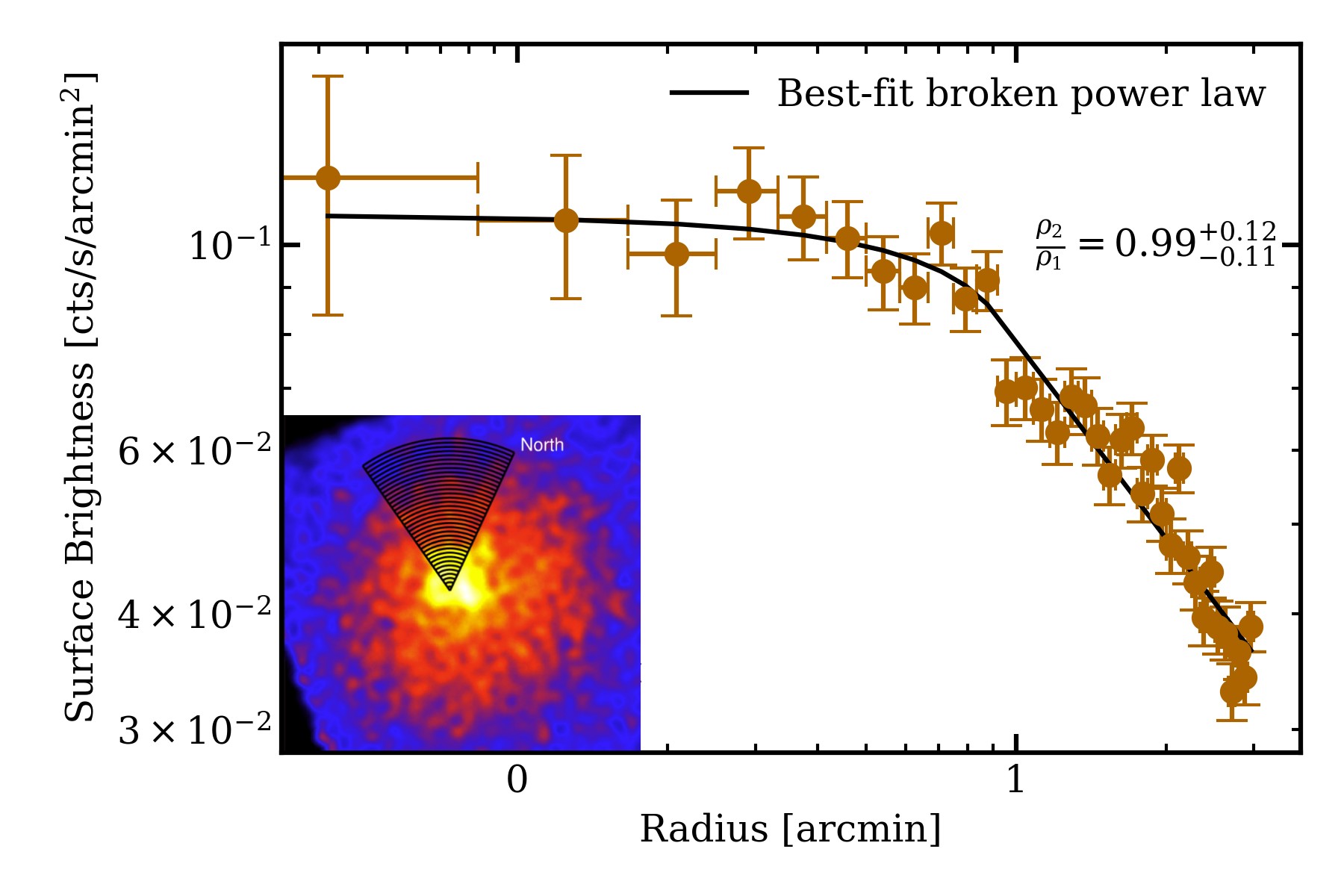}
    \caption{\textit{Chandra} surface brightness profiles and the corresponding \mbox{best-fit} broken power law model of three sectors along the east (top), south (middle), and north (bottom) directions in the 0.5--2.3\thinspace\unit{keV} band. The radial extent of these profiles is $R=\ang{;2}$. The best-fit value of the parameter $J$ of Eq.\thinspace\ref{eq:sbjump} is displayed on each plot, and the inset displays the orientation of the sectors using the 0.5--2.3\thinspace\unit{keV} \textit{Chandra} image.}
    \label{fig:shockfronts}
\end{figure}
For the fit, we used broad, flat priors that allow for a proper exploration of the parameter space without making any assumption about the true value of the parameters. In addition, we restricted $\alpha_1$ and $\alpha_2$ to only positive values to avoid nonphysical slopes for the respective power laws. The best-fit parameters of Eq.\thinspace\ref{eq:sbjump} for the three sectors are given in \cref{tab:chandrafitpara}. In \cref{fig:shockfronts}, we display the \textit{Chandra} surface brightness profiles of the eastern, southern, and northern sectors and their respective best-fit broken power law models. The discontinuity is best sampled in the eastern sector, followed by the southern sector, whereas the northern sector only shows a modest steepening in the surface brightness profile. Regarding the overall shape of the profiles, we observe small variations within $R\approx\ang[angle-symbol-over-decimal]{;0.4}$, followed by the discontinuity at $R\approx\ang{;1}$ in the eastern and southern sectors. In contrast, the northern sector has a flatter profile until $R=\ang{;1}$, after which a sudden decrease in magnitude of $3.6\sigma$ is observed. However, the magnitude of the best-fit density contrast is $\approx$1, which remained consistent with different robustness tests, for example, varying the bin width and improving the S/N. We therefore refrain from interpreting the northern sector as a robust discontinuity. We also note that beyond the discontinuity, the profile transitions from a shallow decline till $R\approx\ang[angle-symbol-over-decimal]{;1.5}$ into a steeper power \mbox{law-like} behavior. A similar profile shape is also observed beyond the discontinuity in the east. Furthermore, we note that the southern discontinuity exhibits the strongest density contrast, albeit with the lowest S/N. In addition, the profile in this sector exhibits the flattest slope of $\alpha_2=0.58\substack{+0.02\\-0.03}$ among all the sectors beyond the discontinuity.\\ 
\indent
The eastern and southern sectors exhibit density contrasts of \mbox{$J=1.28\substack{+0.09\\-0.05}$} and $1.59\substack{+0.15\\-0.07}$, respectively, which are consistent with weak ICM shocks \citep[e.g.,][]{Markevitch2007}. However, this interpretation requires direct spectroscopic measurements of the temperature structure across the discontinuities. Previous studies by \citet{2004hydra,2006xmmhydra} have performed detailed spectral analyses of the cluster core and do not report temperature or metallicity jumps within $R<\ang{;2}$. However, the available data are insufficient to conclusively distinguish between shocks, cold fronts, or other ICM discontinuities. \citet{2024radiohydra} discussed the possibility of these edges being a sloshing feature, but in our GGM filtered and unsharp masked \textit{Chandra} images (\cref{fig:chandraggm+unsharp}), this feature is primarily localized near NGC~3311 and lacks any signs of large-scale spiraling gas flows in its vicinity and closer to NGC 3309. We also note that similar examples of ongoing ram-pressure stripping in galaxy groups, such as NGC~1404 in the Fornax cluster \citep{2005ApJ...621..663M} and NGC~4552 in the Virgo cluster \citep{2006ApJ...644..155M}, exhibit density contrasts at the edges of stripped gas that are similar in magnitude. \\
\indent 
Moreover, the northern part of the Flying Fox spatially overlaps with the eastern and southern discontinuities along the line of sight (\cref{fig:chandra_TGSS}). This suggests that these discontinuities may contribute to diffuse radio emission. Although diffusive shock acceleration \citep[e.g.,][]{1983RPPh...46..973D} is inefficient at weak ICM shocks, an older population of cosmic rays that originated from past turbulence and AGN outbursts could be reaccelerated from these features and emit synchrotron radiation \citep[e.g.,][]{2007ApJ...669..729K,2012ApJ...756...97K}. Such a population could originate from the fossil radio plasma of a previous outburst from the radio-loud AGN of NGC 3311. This interpretation is consistent with the steep average spectral index of $\alpha=-1.4$ reported by \citet{2024radiohydra}. However, they also reported a brightness decrease from 3.2~mJy~beam$^{-1}$ in the southeast to $\approx$0.7~mJy~beam$^{-1}$ in the northwest, which hints at an additional contribution to the Flying Fox from the ram-pressure stripping of NGC~3312 \mbox{\citep[e.g.,][]{2022A&A...668A.184H}} due to its proximity to the southern tail of the radio emission. Therefore, based on our analysis, the most plausible formation scenario for Flying Fox is that the brighter southeastern section of the source is the fossil radio emission from NGC~3311 with an additional contribution from NGC~3312. Furthermore, the eastern and southern discontinuities likely re-inject energy into the fossil plasma closer to NGC~3311 and produce the weaker northwestern section of the Flying Fox.
\subsection{eROSITA normalization, temperature, and metallicity profiles}
\label{sec:norm,temp,met}
\begin{figure}[h]
    \centering   
    \includegraphics[width=1.0\linewidth]{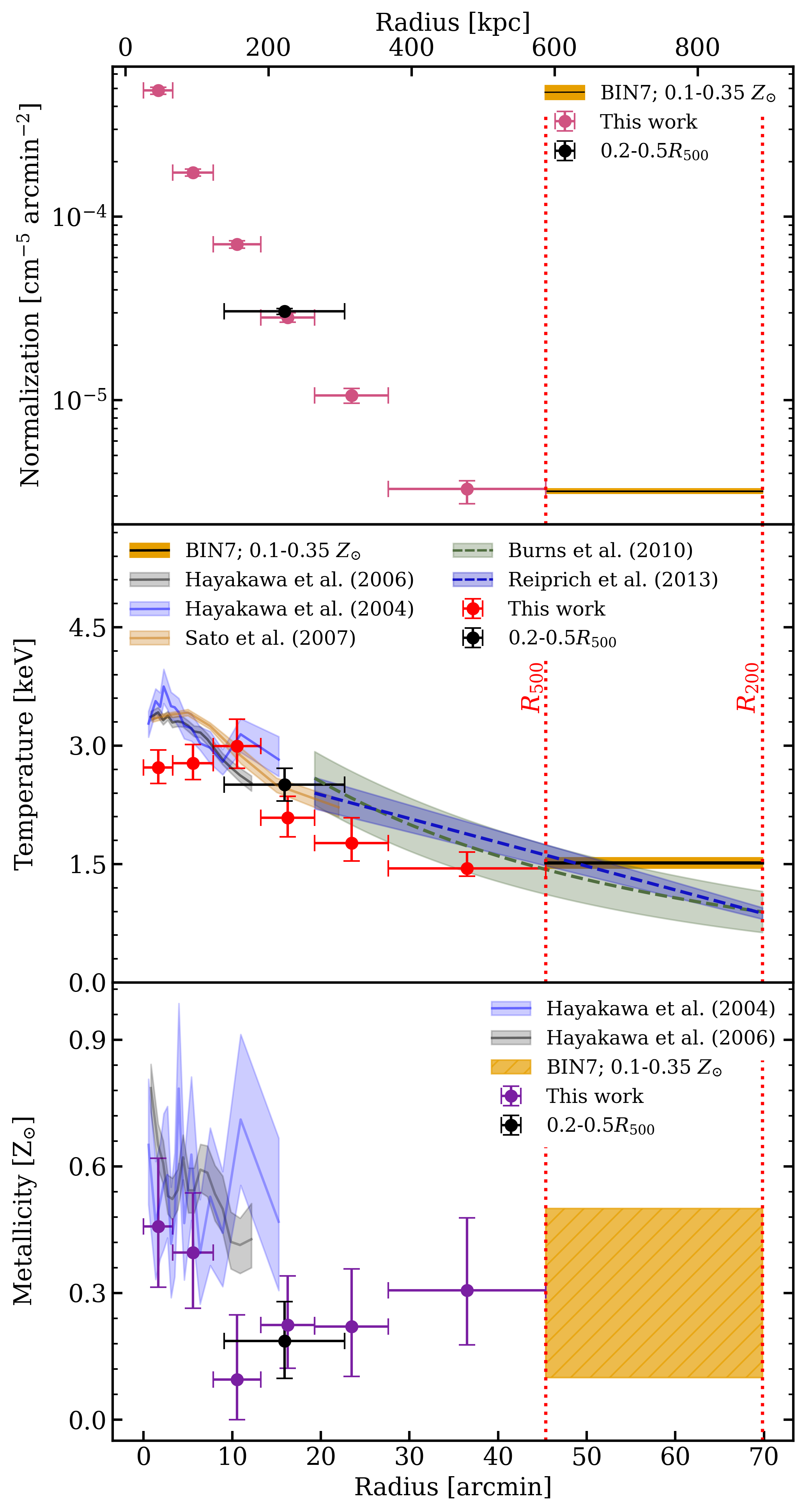}
    \caption{eROSITA normalization (top), temperature (middle), and metallicity (bottom) profiles of Abell 1060 within the radial range $0\leq R \leq R_\mathrm{200}$. The previous estimates of the temperature and metallicity profiles and the expected profiles in the outskirts from \citet{burns} and \citet{2013reiprich} are plotted using solid and dashed lines, respectively. Additionally, their respective $1\sigma$ uncertainties are also plotted using colored shaded regions.}
    \label{fig:TempNormMetprofile}
\end{figure}

For the estimation of the normalization, temperature, and metallicity profiles of Abell 1060, we extracted the cluster spectrum from seven radial annuli within the radial range \mbox{$0\leq R \leq$ \rtwo}. The widths of these annuli are optimized such that the S/N declines by a maximum of $2\sigma$ per bin. Next, we fitted the cluster spectra with the full spectral model (Eq.\thinspace\ref{eq:clustermodel}) following the procedure described in Sect.\thinspace\ref{sec:spectralanalysis}. The best-fit parameters thus obtained for each bin are mentioned in \cref{tab:tempprof} and the normalization, temperature, and metallicity profiles are displayed in \cref{fig:TempNormMetprofile}. For the $R_\mathrm{500}\leq R \leq R_\mathrm{200}$ bin, the spectrum could not be fitted due to poor S/N. Therefore, we fixed the metallicity of the $\texttt{apec}_\mathrm{ICM}$ component to a range of values between $0.1\leq Z \leq 0.5\,Z_{\odot}$ and represented the median and standard deviation of the \mbox{best-fit} normalizations and temperatures using solid black lines and orange shaded regions, respectively, in \cref{fig:TempNormMetprofile}. However, the results from this bin should be interpreted with caution given their dependence on the assumed metallicity range.\\
\indent
In \cref{fig:TempNormMetprofile} (top), we display the radial distribution of the \texttt{apec}$_\mathrm{\texttt{ICM}}$ normalization per unit area until \rtwo. The \texttt{apec} normalization is described as 
\begin{equation}
\label{eq:apecnorm}
    \mathrm{Normalization}=\frac{10^{-14}}{4\pi[D_\mathrm{A}(1+z)]^2}\;\mathrm{EM},
\end{equation}
where $D_{\text{A}}$ is the angular diameter distance in cm and EM is the emission measure, which depends on the electron number density ($n_\mathrm{e}$) as 
\begin{equation}
\label{eq:EM}
   \mathrm{EM}=\int n_\mathrm{e}^2\;\mathrm{d}V,
\end{equation}
where $\text{d}V$ is the volume element in cm$^{3}$. We observe the highest normalization in the central bin, followed by a steady decline by two orders of magnitude until \rtwo\;in the profile. This is because the \texttt{apec} normalization has an $n_\mathrm{e}^2$ dependence (Eqs. \ref{eq:apecnorm} and \ref{eq:EM}) and Abell 1060 lacks substructures (Sect.\thinspace\ref{sec:xrayimages}).
Next, we present the eROSITA temperature profile of Abell~1060 in \cref{fig:TempNormMetprofile} (middle). We observe a peak temperature of \mbox{$k_\mathrm{B}T=3.00\substack{+0.34\\ -0.28}$\thinspace\unit{keV}} from the third bin, followed by a $\approx$$50\%$ decline at \rfive. Between \rfive\,and \rtwo, our temperature estimate remains within $0.4\sigma$ of the previous bin. This decline is consistent with the decline in the temperature profile and its intrinsic scatter of 10-15\% at \rfive\;reported by \citet{2019A&A...621A..41G}. We also observe a weak inward decline in temperature within \mbox{$R<\ang{;10}$}. Furthermore, we obtained an average ICM temperature of \mbox{$\langle k_\mathrm{B}T\rangle=2.51\substack{+0.21\\-0.21}$\thinspace\unit{keV}} from the 0.2-0.5\rfive\;annulus, whereas previous ASCA, \textit{Chandra}, \textit{XMM-Newton}, and \textit{Suzaku} profiles \citep[e.g.,][]{2000hydra,2001ascahydra,2004hydra,2006xmmhydra,2007hydra} unanimously reported an average temperature of $\langle k_\mathrm{B}T\rangle\gtrsim3.1$\thinspace\unit{keV}. This variation in ICM temperatures is likely the combined effect of the following: 1) \textit{Chandra}, \textit{XMM-Newton}, and \textit{Suzaku} have a harder energy response compared to eROSITA, and 2) the ICM has a multitemperature structure, and the cooler, high EM component dominates the emission-weighted best-fit temperature estimate \citep{2004Mazzotta}. Within $R\approx\ang{;8}$, our profile is consistent within the cross-calibration limits between \textit{Chandra}, \textit{XMM-Newton}, \textit{Suzaku}, and eROSITA \citep[e.g.,][]{2013A&A...552A..47K,calibration}. However, in the radial range $\ang{;8}\leq R \leq \ang{;27}$, the previous temperature profiles are unexpectedly consistent with our profile. We also compared our profile in the radial range \mbox{0.43\rfive\thinspace$\leq R\leq\;$\rfive}\;to the average temperature profile obtained from hydrodynamical simulations by \citet{burns} and 162 \textit{Suzaku} temperature measurements from \citet{2013reiprich}, both of which are normalized using the average temperature of \mbox{$\langle k_\mathrm{B}T\rangle=2.51\substack{+0.21\\-0.21}$\thinspace\unit{keV}} (dashed green and blue lines and shaded regions in \cref{fig:TempNormMetprofile} middle). We note that our profile is broadly consistent with the average profiles within error bars.\\
\indent
From \cref{fig:TempNormMetprofile} (bottom), we note that the best-fit metallicity profile of Abell 1060 is relatively flat. We observe a nearly constant metallicity of $Z\approx0.43\thinspace Z_\odot$ until $R=\ang[angle-symbol-over-decimal]{;7.85}$ followed by a drop of $\approx$$1\sigma$ in the $\ang[angle-symbol-over-decimal]{;7.85}\leq R\leq \ang[angle-symbol-over-decimal]{;13.20}$ bin, which is statistically not significant. The profile then stays flat until \rfive\;with nearly a constant metallicity of $Z\approx0.25\thinspace Z_{\odot}$. We also estimated an average metallicity of $\langle Z \rangle=0.19 \substack{+0.10\\-0.10}\; Z_\odot$ from the 0.2-0.5\rfive\;annulus. Furthermore, we compared our metallicity profile with \textit{Chandra} and \textit{XMM-Newton} profiles from \citet{2004hydra,2006xmmhydra} by applying a correction factor of 1.48 to them, to correct for the difference in solar abundance tables \citep[e.g.,][]{angiecent}. This factor is the ratio of the assumed solar Fe/H values in \citet{1989GeCoA..53..197A} and \citet{asplund}. In \cref{fig:TempNormMetprofile} (bottom), both these profiles, depicted by the blue and grey shaded regions, are in agreement with our profile within error bars until $R= \ang[angle-symbol-over-decimal]{;7.85}$. In general, the \textit{Chandra} profile fluctuates around an average metallicity of $Z=0.59\thinspace Z_\odot$, whereas the \textit{XMM-Newton} profile shows a more expected decline from $Z=0.79\thinspace Z_\odot$ in the innermost bin to $Z=0.43\thinspace Z_\odot$ at $R= \ang[angle-symbol-over-decimal]{;12.16}$.  \\
\indent
Furthermore, based on our imaging and spectral analyses, Abell~1060 shows intermediate CC and NCC cluster properties. The shallow inward decline within $R=\ang[angle-symbol-over-decimal]{;13.2}$ in our temperature profile, as well as the weak central surface brightness cusp, are initial signs of radiative cooling, which is indicative of a transitionary state known as a WCC \citep{2010A&A...513A..37H}. Moreover, the central cooling time range of the ICM within $R<\ang{;5}$ in the ACCEPT sample \citep{2009ApJS..182...12C} is between 3.69 and 5.33 Gyr, which is typical of WCC clusters. This timescale is also consistent with the last major merger period estimated by \citet{2007hydra} of $\sim$3 Gyr. Furthermore, \citet{2010A&A...513A..37H} also classified Abell 1060 as a WCC cluster based on the cooling time in their CC diagnostics of the HIFLUGCS clusters. On the other hand, \citet{2006xmmhydra} reported small-scale temperature enhancements at $R\approx\ang{;7}$ in the southeast direction that originated less than a gigayear ago. However, we do not detect any signs of that feature in our eROSITA surface brightness or temperature profiles.

\section{Summary and conclusion}
\label{sec:conclusion}
In this work, we performed a thorough imaging and spectral analysis using eRASS:4 data with a FoV of $\approx$300 square degrees centered on Abell 1060 that included a highly varying and complex CXB structure. Moreover, we used the archival \textit{Chandra} observation of Abell 1060 to probe the ICM within $R=\ang{;4}$. We summarize our results from these analyses below: 
\begin{enumerate}
    \item[--]  We determined that Abell 1060 has a relaxed ICM morphology within \rfive\;from the 0.2--2.3\thinspace\unit{keV} eROSITA image. The ICM is spherically symmetric and lacks any major substructures. The X-ray peak coincides with NGC~3311 along the line of sight (Sect.\thinspace\ref{sec:xrayimages}).
    \item[--] We analyzed the NED spectroscopic redshift distribution of Abell 1060 within \rtwo\;and obtained a \mbox{best-fit} \mbox{$\mu_\mathrm{spec}\pm\sigma_\mathrm{spec}=0.0121\pm0.0027$} via a Gaussian fit. We also obtained nearly identical values from an adaptive \mbox{$n\sigma$-clipping} algorithm. The redshift distribution is unimodal and the peak in the 2D galaxy distribution displays high spatial correlation with the X-ray peak (Sect.\thinspace\ref{sec:galdistrib}).
    \item[--] We did not find a galaxy overdensity directly between Abell~1060 and the Antlia cluster in the redshift range \mbox{$0\leq z \leq 0.03$}. In the redshift range $0.03\leq z \leq 0.08$, several indirect overdensities that connect the two clusters are apparent in the NED and 2MASS maps (\cref{fig:2MASS_wf}). However, the Antlia SNR obscures any possible filamentary X-ray emission present there (Sects. \ref{sec:xrayimages} and \ref{sec:galdistrib}).
    \item[--] The modified $\beta$-model (Eq.\thinspace\ref{eq:modbetamodel}) is a good fit ($\chi_\mathrm{red}^2=0.9$) to the 0.2--2.3\thinspace\unit{keV} eROSITA surface brightness profile until 3\rtwo\;and successfully models the weak central surface brightness cusp. The best-fit $\beta$ and $r_\mathrm{c}$ parameters are consistent with previous estimates from \citet{2004hydra} and \citet{2007A&A...466..805C}. 
    \item[--] The outskirts of Abell 1060 are actively accreting baryons, as evidenced by the detection of soft X-ray excesses in the north and northwest beyond \rtwo, both of which strongly correlate with the 2D galaxy distribution. The northern and northwestern excesses reach peak fiducial significances of 3.9$\sigma$ and 3.3$\sigma$, respectively, and remain detectable out to 1.4\rtwo\;and 1.1\rtwo\;with median significances of 3$\sigma$ and 2.4$\sigma$. We also identify a tentative southwestern excess with a peak fiducial significance of 2.9$\sigma$; however, the foreground emission associated with region C makes its true extent difficult to determine (Sect.\thinspace\ref{sec:sectorsb}).
    \item[--] We discovered two surface brightness discontinuities at the eastern and southern edges of the ram-pressure stripped halo of NGC 3311 with density contrasts of \mbox{$J=1.28\substack{+0.09\\-0.05}$} and $1.59\substack{+0.15\\-0.07}$, respectively. These features, together with NGC~3312 and fossil radio emission from NGC~3311, likely contribute to the Flying Fox (Sect. \thinspace \ref{sec:sbshock}).
    \item[--] We estimated an average ICM temperature and metallicity of $\langle k_\mathrm{B}T\rangle=2.51\substack{+0.21\\-0.21}$\thinspace\unit{keV} and $\langle Z \rangle=0.19\substack{+0.10\\-0.10}\thinspace Z_\odot$, respectively from the 0.2-0.5\rfive\;annulus. The temperature profile is broadly consistent with the average profiles from \citet{burns} and \citet{2013reiprich} in the radial range 0.43\rfive\thinspace$\leq R \leq$ \rfive.
    \item[--] We classified Abell 1060 as a WCC cluster based on our observations of the shallow inward decline in the temperature profile within $R=\ang[angle-symbol-over-decimal]{;13.2}$, weak surface brightness cusp, no current signs of AGN activity, relaxed morphology within \rfive, and a central cooling time between 3.69-5.33~Gyr in the ACCEPT sample \citep{2009ApJS..182...12C}. This behavior is likely caused by the initiation of radiative cooling in the core following the last major merger event.     
\end{enumerate}
Through our employed combination of eROSITA and \textit{Chandra} observations in this work, we significantly expanded the current understanding of the physical properties of Abell~1060’s ICM for the first time beyond \rfive. Furthermore, the correlation between X-ray features and multiwavelength data from TGSS, 2MASS, and NED showed nonthermal emission, a unique dynamical state, and the surrounding large-scale structure. However, future observations of the central $R=\ang{;10}$ of the cluster using LOFAR \citep{2013A&A...556A...2V} and the high X-ray spectral resolution of XRISM \citep{xrism} will help us better understand the ICM kinematics and radio emissions from NGC~3309, NGC~3311, and NGC~3312. Furthermore, observations of potential X-ray filamentary emission using HUBS \citep{2023SCPMA..6699513B} along the optical/IR galaxy overdensities between Abell 1060 and the Antlia cluster will be useful to map the WHIM filaments closer to 3\rtwo.

\begin{acknowledgements}
    We thank the anonymous referee for their comments on this manuscript. This work is partially funded by the Deutsche Forschungsgemeinschaft (DFG, German Research Foundation) under Germany’s Excellence Strategy – EXC 3037 – 533607693.  JD acknowledges funding by the Federal Ministry of Education and Research (BMBF) and the Ministry of Culture and Science of the State of North Rhine-Westphalia (MWK) as part of TRA Matter and the Excellence Strategy of the federal and state governments. This work is based on data from eROSITA, the soft X-ray instrument aboard SRG, a joint Russian-German science mission supported by the Russian Space Agency (Roskosmos), in the interests of the Russian Academy of Sciences represented by its Space Research Institute (IKI), and the Deutsches Zentrum für Luft- und Raumfahrt (DLR). The SRG spacecraft was built by Lavochkin Association (NPOL) and its subcontractors, and is operated by NPOL with support from the Max Planck Institute for Extraterrestrial Physics (MPE). The development and construction of the eROSITA X-ray instrument was led by MPE, with contributions from the Dr. Karl Remeis Observatory Bamberg \& ECAP (FAU Erlangen-Nuernberg), the University of Hamburg Observatory, the Leibniz Institute for Astrophysics Potsdam (AIP), and the Institute for Astronomy and Astrophysics of the University of Tübingen, with the support of DLR and the Max Planck Society. The Argelander Institute for Astronomy of the University of Bonn and the Ludwig Maximilians Universität Munich also participated in the science preparation for eROSITA. The eROSITA data shown here were processed using the eSASS software system developed by the German eROSITA consortium. This research has made use of data obtained from the \textit{Chandra} Data Archive provided by the \textit{Chandra} X-ray Center (CXC). We thank the staff of the GMRT that made these observations possible. GMRT is run by the National Centre for Radio Astrophysics of the Tata Institute of Fundamental Research. The Second Palomar Observatory Sky Survey (POSS-II) was made by the California Institute of Technology with funds from the National Science Foundation, the National Aeronautics and Space Administration, the National Geographic Society, the Sloan Foundation, the Samuel Oschin Foundation, and the Eastman Kodak Corporation. This publication makes use of data products from the Two Micron All Sky Survey, which is a joint project of the University of Massachusetts and the Infrared Processing and Analysis Center/California Institute of Technology, funded by the National Aeronautics and Space Administration and the National Science Foundation. This research has made use of the NASA/IPAC Extragalactic Database, which is funded by the National Aeronautics and Space Administration and operated by the California Institute of Technology.
\end{acknowledgements}

\bibliographystyle{aa}
\bibliography{references}
\begin{appendix}
\nolinenumbers
\section{$N_\mathrm{H,tot}$ map}
The $N_\mathrm{H,tot}$ map is prepared following the expression from \citet{2013willy}, which is given as
\begin{equation}
    N_\mathrm{H,tot}=N_\mathrm{HI}+2N_\mathrm{H_2}.
\end{equation}
\begin{figure}[H]
    \centering
    \includegraphics[width=0.95\linewidth]{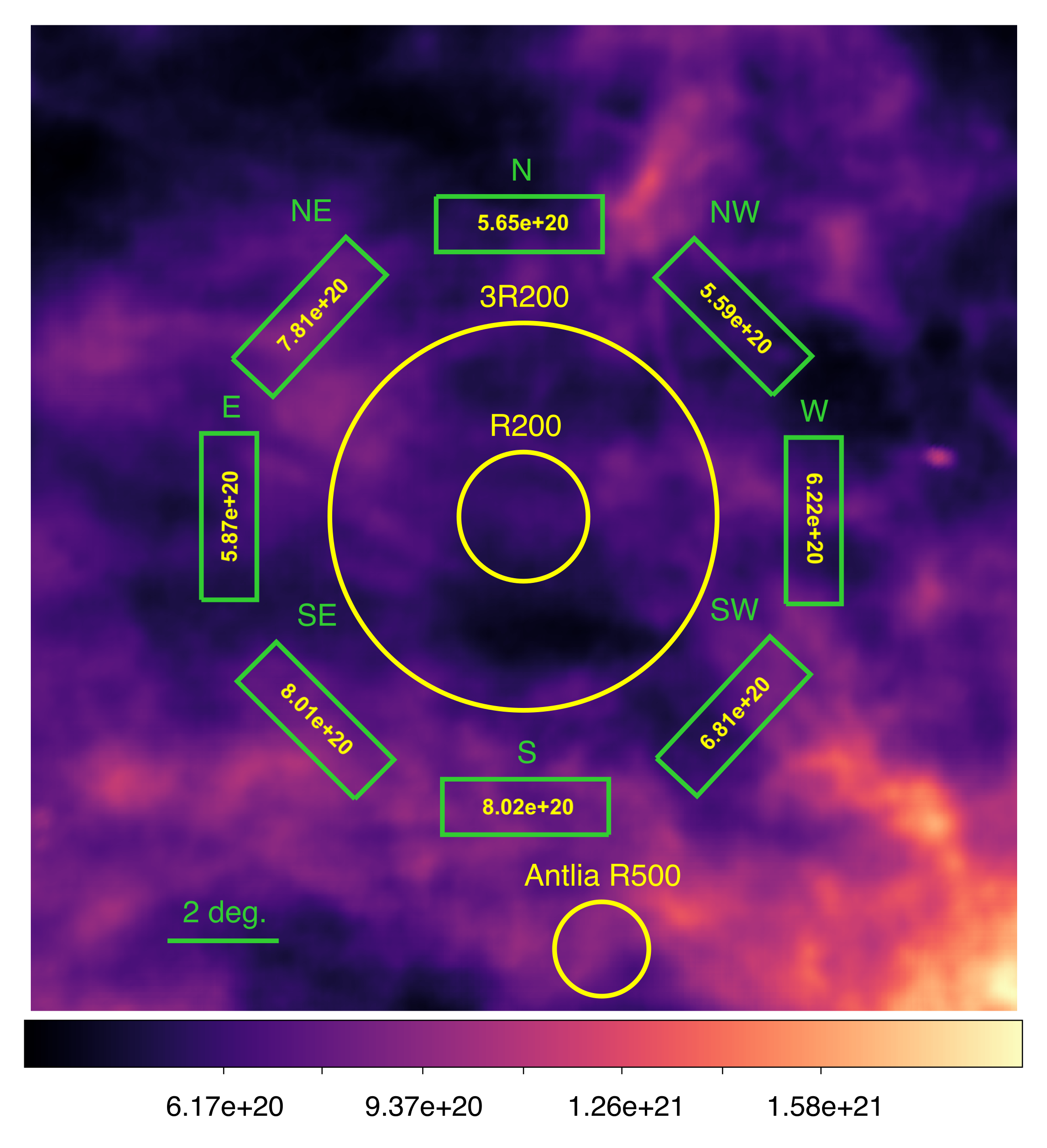}
    \caption{$N_\mathrm{H,tot}$ map (unit is cm$^{-2}$) displaying the spatial variation of the total hydrogen column density in our FoV with respect to the characteristic radii of Abell 1060 and the Antlia cluster. The background boxes and their respective median $N_\mathrm{H,tot}$ are annotated.}
    \label{fig:nhtotmap}
\end{figure}

\section{X-ray images}
\label{app:xrayimages}
\begin{figure}[H]
    \centering
    \includegraphics[width=0.9\linewidth]{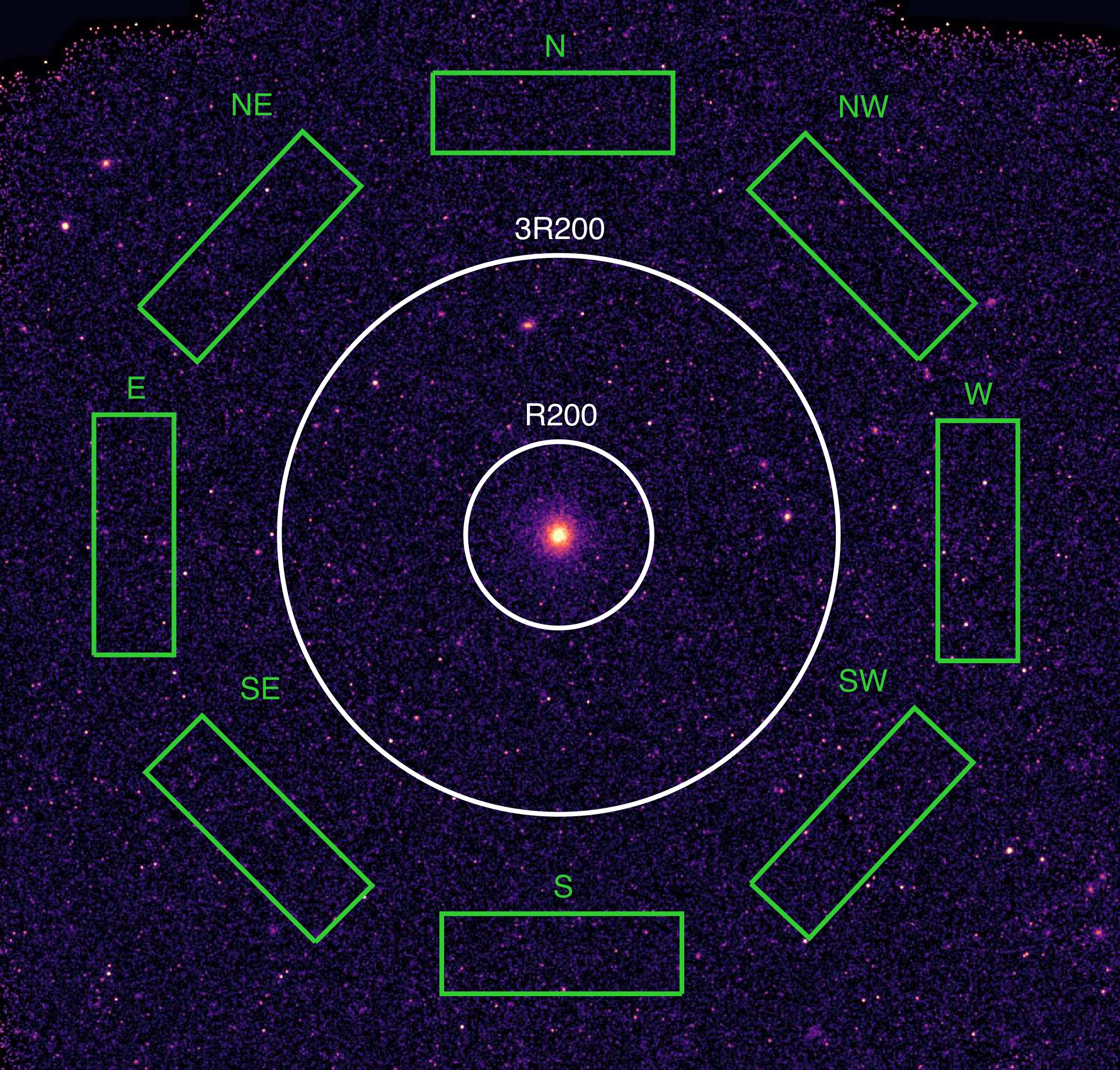}
    \caption{Fully corrected eROSITA image in the 0.2--2.3\thinspace\unit{keV} band, smoothed using a Gaussian kernel of 18 pixels width. The eight background regions used to estimate the CXB for the surface brightness and spectral analyses are displayed using green boxes.}
    \label{fig:cxbsetup}
\end{figure}
\begin{figure}[H]
    \centering
    \includegraphics[width=1.0\linewidth]{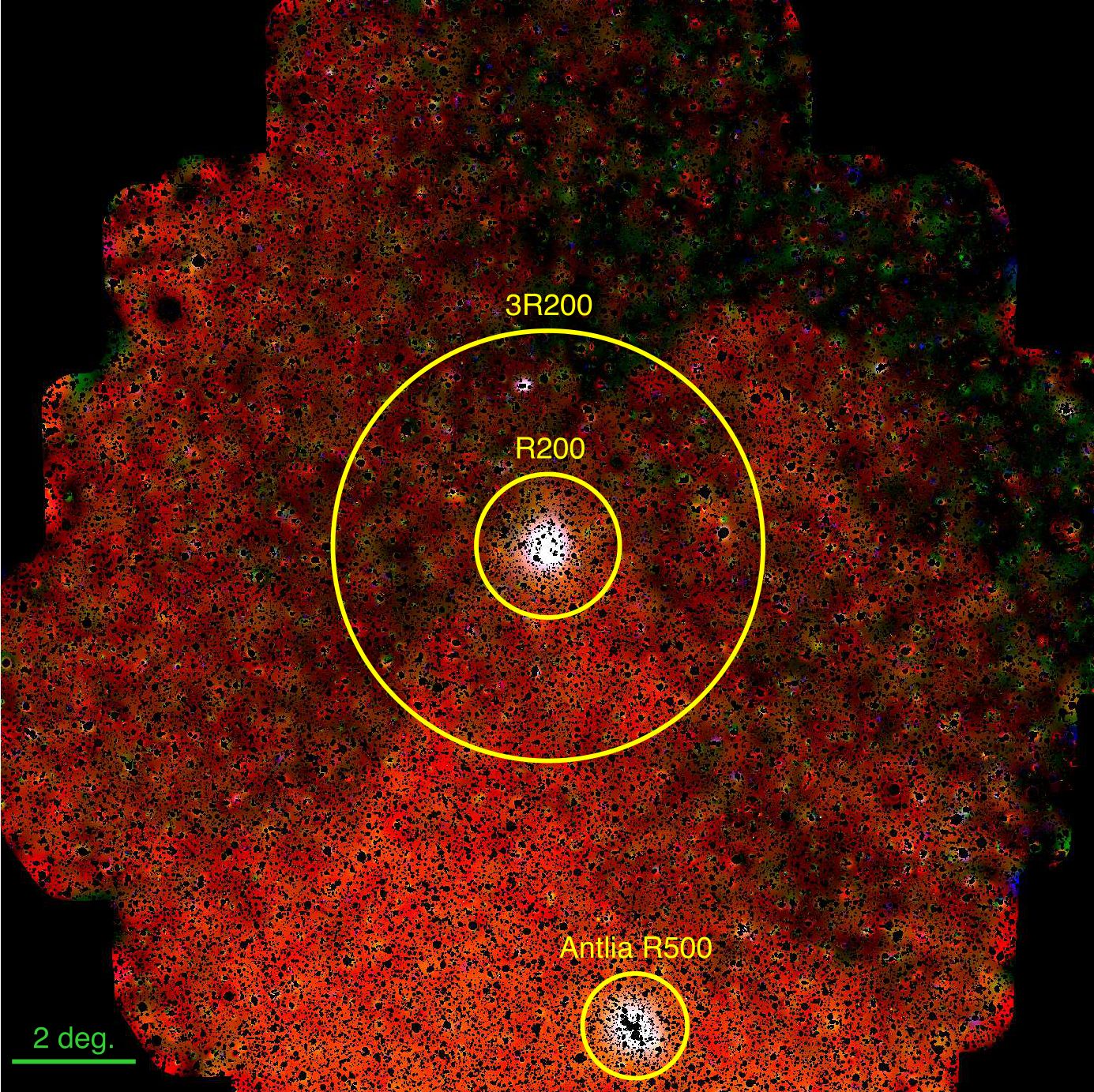}
    \caption{Point source removed eROSITA wavelet filtered RGB image, prepared using the energy bands 0.2--0.8\thinspace\unit{keV} (red), 0.8--1.2\thinspace\unit{keV} (green), and 1.2--2.3\thinspace\unit{keV} (blue).}
    \label{fig:wfrgb}
\end{figure}
\begin{figure}[H]
    \centering
    \includegraphics[width=1.0\linewidth]{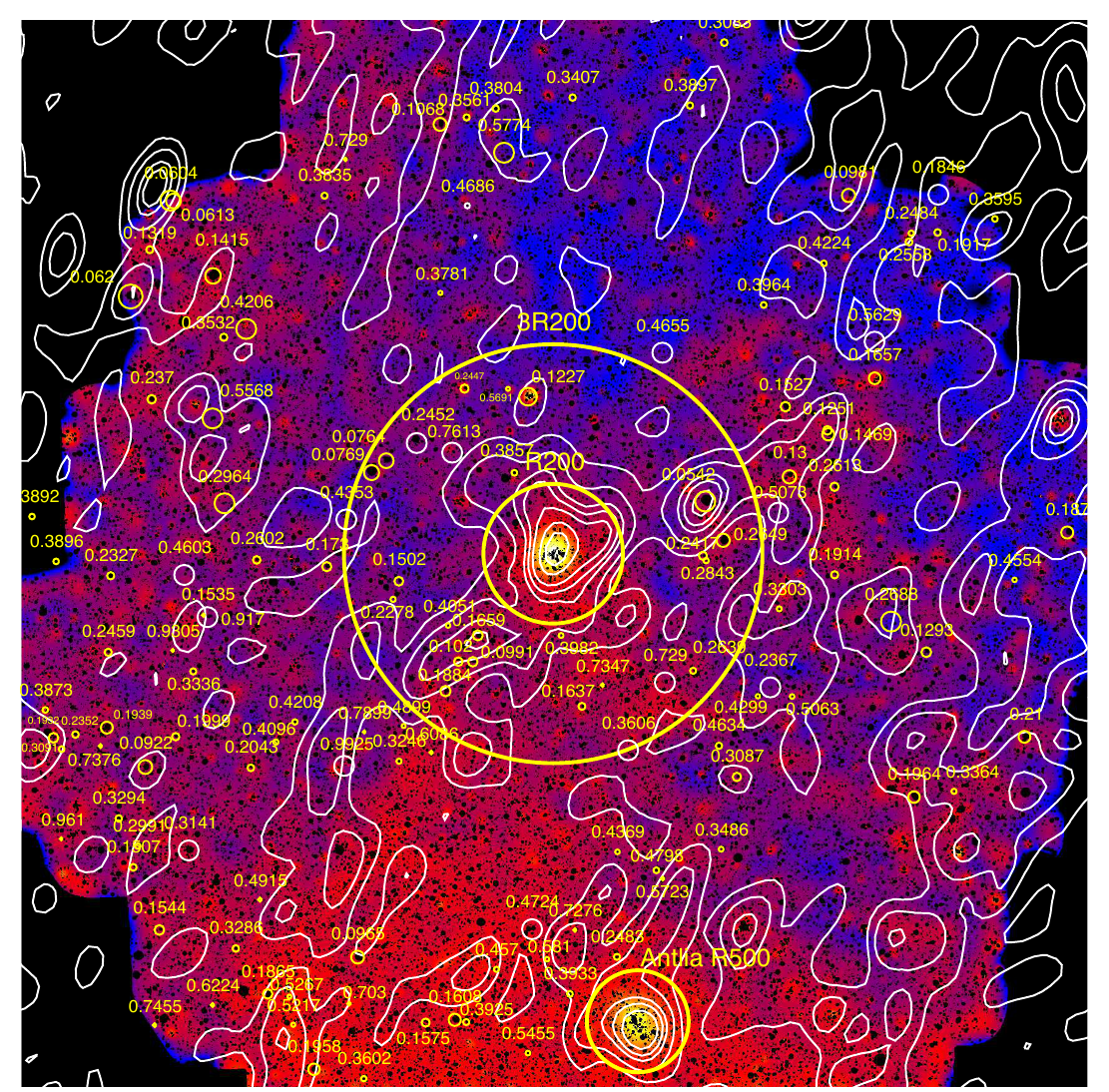}
    \caption{Same as \cref{fig:fc_wf} but overlaid with 2MASS galaxy distribution contours and eRASS:1 background clusters. The cluster positions are plotted in yellow (the circle represents the \rfive) and white (unknown \rfive, marked via circles of $R=\ang{;10}$).}
    \label{fig:2MASS_wf}
\end{figure}

\section{X-ray surface brightness analysis}
\label{app:sbanalysis}
\begin{figure*}[t]
    \centering
    \includegraphics[width=0.9\linewidth]{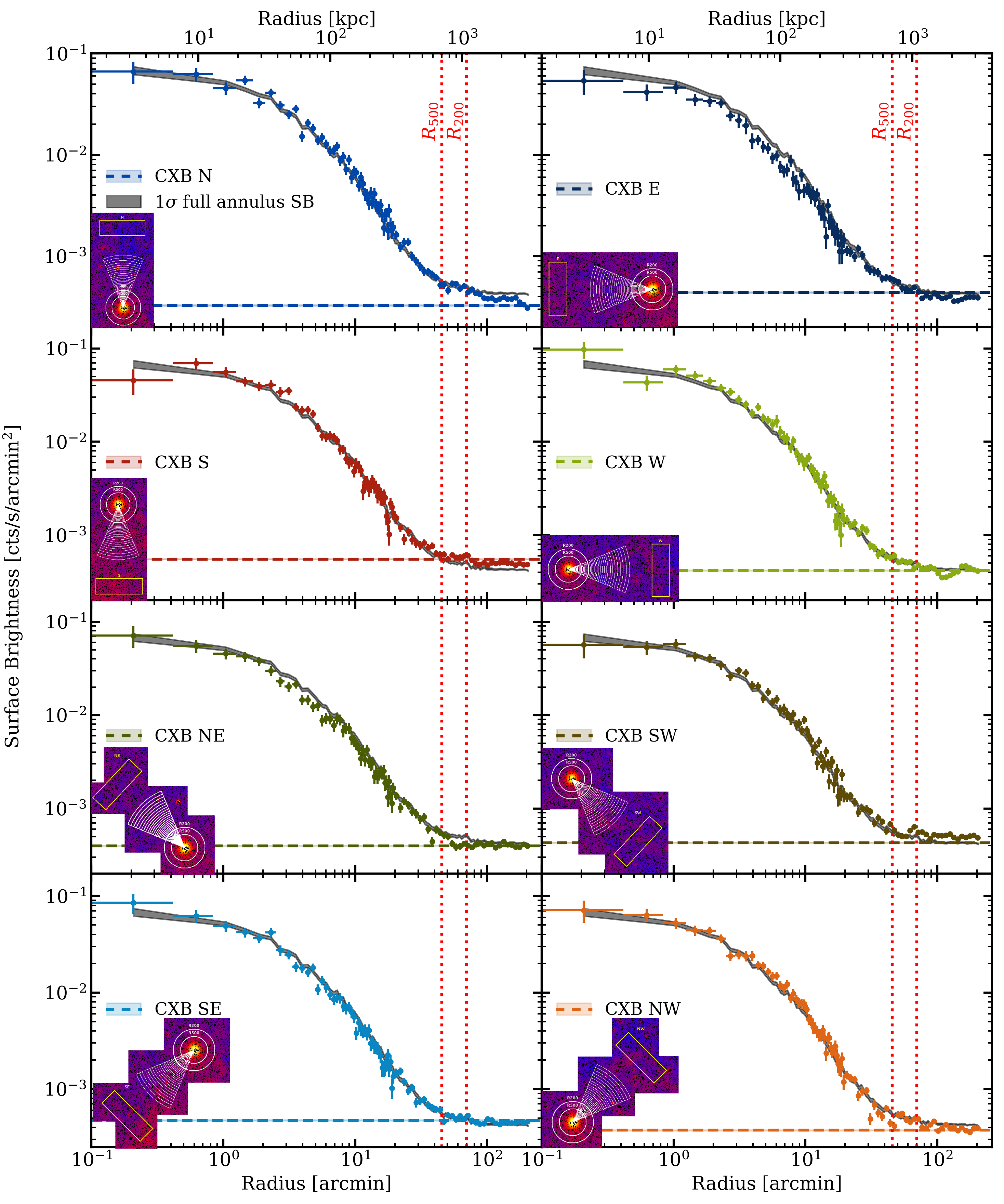}
    \caption{eROSITA sector surface brightness profiles in the 0.2--2.3\thinspace\unit{keV} band. The gray shaded region depicts the $1\sigma$ error bars of the full annulus profile. The inset in each sector plot gives the orientation of the sector and the background box used to estimate the CXB level.}
    \label{fig:sectorsb}
\end{figure*}

\begin{table*}[t]
   \centering
   \begin{threeparttable}[b]
   \caption{Best-fit parameters of the broken power law model (Eq.\thinspace\ref{eq:sbjump}) for each sector.}
   \label{tab:chandrafitpara}
   \begin{tabular}{@{}ccccccc@{}}
   \toprule
   Sector  & $\alpha_1$ & $\alpha_2$ & $r_{\text{f}}^{(a)}$ & $S_\mathrm{X}(0)^{(b)}$ & $J$ & $\chi^2 /\text{d.o.f.}$  \\ \midrule
   East &$0.05\substack{+0.06 \\ -0.03}$&$0.76\substack{+0.02 \\ -0.02}$&$0.82\substack{+0.06 \\ -0.04}$&$-1.20\substack{+0.04 \\ -0.05}$& $1.28\substack{+0.08\\-0.05}$&48.26/41.00\\
   South &$0.03\substack{+0.03 \\ -0.02}$&$0.58\substack{+0.02 \\ -0.03}$&$0.67\substack{+0.07 \\ -0.06}$&$-1.22\substack{+0.08 \\ -0.07}$&$1.59\substack{+0.15 \\ -0.07}$& 16.94/10.00\\
   North &$0.05\substack{+0.20 \\ -0.25}$&$0.85\substack{+0.02 \\ -0.02}$&$0.88\substack{+0.01 \\ -0.01}$&$-1.31\substack{+0.09 \\ -0.09}$&$0.99\substack{+0.12 \\ -0.11}$&51.20/31.00 \\
   \bottomrule
   \end{tabular}
   \tablefoot{
      \tablefoottext{a}{Unit is arcmin.}
      \tablefoottext{b}{Units are counts s$^{-1}$ arcmin$^{-2}$; values are in log base 10.}
   }
   \end{threeparttable}
\end{table*}

We performed the surface brightness analysis of Abell 1060 using eROSITA and \textit{Chandra} to study its brightness distribution and to characterize the morphology of its ICM. We used concentric radial annuli for the profile with a constant binning of \ang{;;25} (approximate width of the eROSITA PSF) in the radial range $0\leq R\leq R_{\mathrm{2500}}$, followed by logarithmic bins until 3\rtwo. For the extraction, we used the eROSITA TM0 photon image, the $N_\text{H,tot}$ corrected exposure map (single TM version), and the PIB map in the 0.2--2.3\thinspace\unit{keV} band. Similarly, for \textit{Chandra}, the cleaned image, the exposure map (in seconds),
and the background map in the 0.5--2.3\thinspace\unit{keV} are used. We used the \texttt{FTOOLS}\footnote{\href{https://heasarc.gsfc.nasa.gov/ftools/}{https://heasarc.gsfc.nasa.gov/ftools/}} task \texttt{funcnts} to calculate the total counts ($\Sigma$\,cts), the statistical uncertainty ($\sqrt{\Sigma\,\text{cts}}$), and the area within a particular region. We then used the expression
\begin{equation}
    \label{eq:sbformula}
    \text{Surface Brightness} = \frac{\Sigma\,\text{cts}_{\text{photon}}-\Sigma\,\text{cts}_{\text{PIB}}}{\overline{t_{\text{exp}}}\cdot A},
\end{equation}
where $\Sigma\,\text{cts}_{\text{photon}}$ and $\Sigma\,\text{cts}_{\text{PIB}}$ are the total counts within a region 
in the X-ray image and the PIB image, respectively, $\overline{t_{\text{exp}}}$ is the average exposure time per pixel, i.e.,~the total 
exposure counts in the region normalized by the total number of pixels, and $A$ is the area of the region in units 
of $\text{arcmin}^2$. The uncertainty on a surface brightness measurement is estimated by the Gaussian propagation of 
the statistical uncertainties on $\Sigma\;\text{cts}_{\text{photon}}$ and $\Sigma\;\text{cts}_{\text{PIB}}$.\\
\indent
In the outskirts, we estimated the statistical significance (in the unit of $\sigma$) of a surface brightness value extracted from a region ($\text{SB}_{\text{region}}$) as compared to the average CXB level ($\text{SB}_{\text{CXB}}$) using the expression 
\begin{equation}
    \label{eq:sbsignif}
    \text{Significance} = \frac{\text{SB}_{\text{region}}-\text{SB}_{\text{CXB}}}{\sqrt{\sigma^2 _{\text{SB,region}} + 
    \sigma^2 _{\text{SB,CXB}}}},
\end{equation}
where $\sigma_{\text{SB,region}}$ and $\sigma_{\text{SB,CXB}}$ are the uncertainties on $\text{SB}_{\text{region}}$ and 
$\text{SB}_{\text{CXB}}$, respectively \citep[e.g.,][]{angiecent}. Similarly, we calculated the residual (in the unit of $\sigma$) between the surface brightness profiles and their respective best-fit models (\cref{fig:SBprofile}) for each radial bin using the expression 
\begin{equation}
    \mathrm{Residual}=\frac{\mathrm{SB}_\mathrm{bin}-\mathrm{Model}_\mathrm{bin}}{\sqrt{\sigma_\mathrm{SB}^2 + \sigma_\mathrm{Model}^2 }},
\end{equation}
where $\mathrm{SB}_\mathrm{bin}$ and $\mathrm{Model}_\mathrm{bin}$ are the profile and model values for a particular radial bin, respectively, and $\sigma_\mathrm{SB}$ and $\sigma_\mathrm{Model}$ are their respective uncertainties. \\
\indent
In \cref{fig:sectorsb}, we display the sector surface brightness profiles in the \mbox{0.2--2.3\thinspace\unit{keV}} band. Within $R=\ang{;10}$, the northeastern, northwestern, southwestern, and southeastern sectors are largely consistent with the full annulus profiles. On the other hand, the northern, eastern, southern, and western sectors show distinct variations. Specifically, the eastern sector shows lower surface brightness with a mean and maximum negative deviation of 1.8$\sigma$ and $3.2\sigma$, respectively, from the full annulus profile. On the contrary, the western sector shows primarily higher surface brightness, albeit of a smaller magnitude than the eastern sector, within the same radial range. In the north and south, the profiles fluctuate around the full annulus profile, but their decline is consistent with its overall shape. Additionally, we observe a faster decline in surface brightness of 2.9$\sigma$ magnitude in the northeast between $\ang{;2}\leq R \leq\ang{;8}$. In addition, we display the best-fit parameters from the surface brightness analysis of the edges of the stripped halo of NGC~3311 in \cref{tab:chandrafitpara}.

\section{CXB spectral analysis}
\label{sec:cxbanalysis}
\begin{figure}[H]
    \centering
    \includegraphics[width=1.0\linewidth]{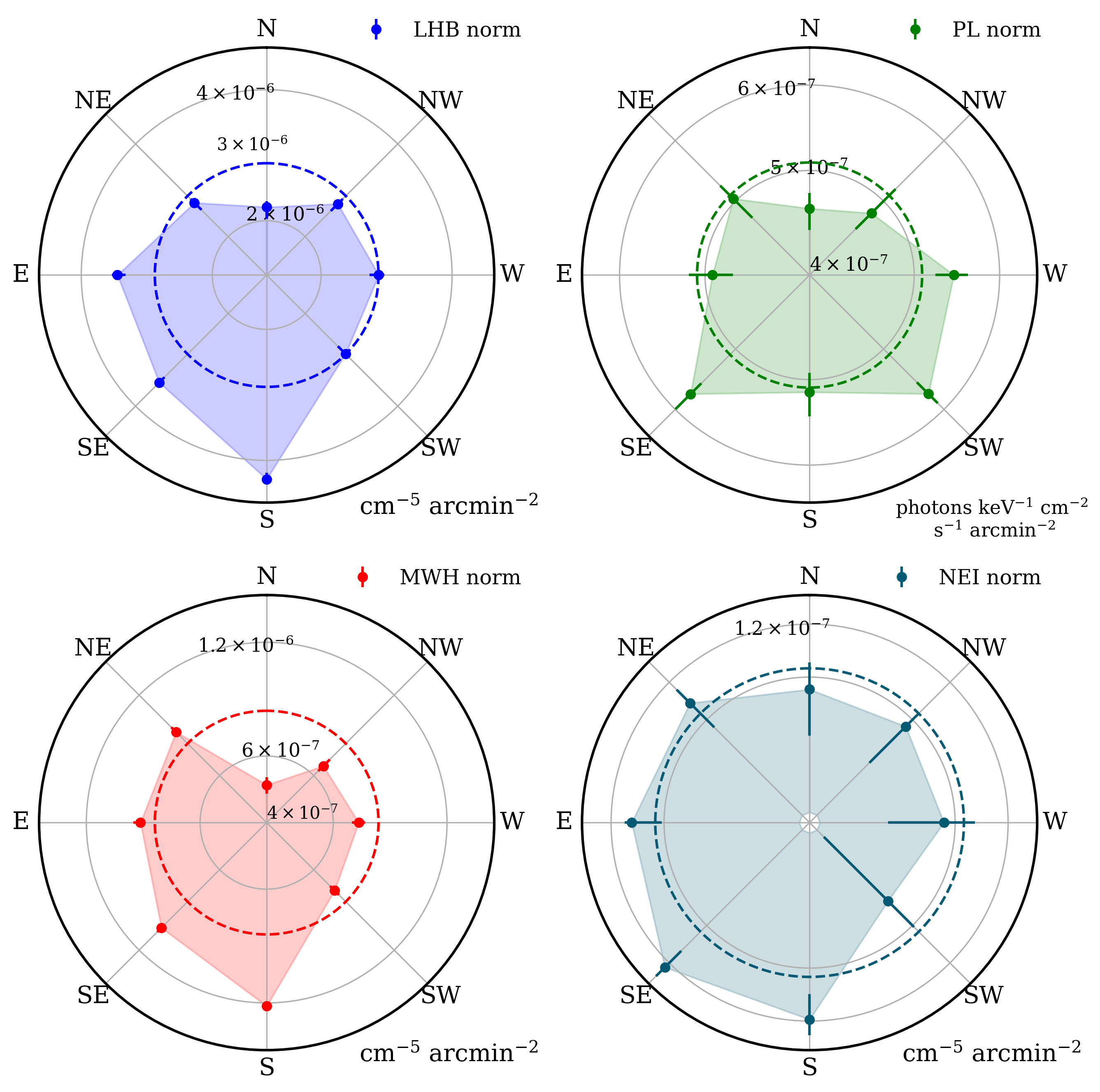}
    \caption{Best-fit normalizations per unit area of the \texttt{apec}$_\mathrm{\texttt{LHB}}$ (top left), \texttt{apec}$_\mathrm{\texttt{MWH}}$ (bottom left), \texttt{nei} (bottom right), and \texttt{powerlaw} (top right) components obtained by fitting the background spectra from the eight background regions shown in \cref{fig:nhtotmap}. The dashed colored lines represent the median of all the regions. The units of the normalizations of each component are mentioned in the bottom right corner of all the subplots.}
    \label{fig:cxbnorms}
\end{figure}

The CXB in the vicinity of Abell 1060 is highly complex and includes various foreground substructures and emission from the Antlia SNR (Sect.\thinspace\ref{sec:xrayimages}). Therefore, we performed a CXB spectral analysis of the eight background spectra (\cref{fig:nhtotmap}) to analyze the variations in the CXB and select a suitable background region for the ICM spectral analysis. We first fitted the CXB spectra with the CXB model from \citet{2024veronica}, which is the same as our CXB model (Eq.\thinspace\ref{eq:clustermodel}) but does not include the \texttt{nei} component. This was done to avoid the observed degeneracy between the LHB and \texttt{nei} components during a previous spectral modeling attempt. For the fit, the temperatures of the \texttt{apec}$_\mathrm{\texttt{LHB}}$ and \texttt{apec}$_\mathrm{\texttt{MWH}}$ are fixed to 0.1\thinspace\unit{keV} and 0.25\thinspace\unit{keV}, respectively \citep{lhb,mwh} and the spectral index of the \texttt{powerlaw} component is fixed to $\Gamma=1.46$ \citep{powind}. Additionally, we assumed solar metallicity and set the redshifts to zero for all the CXB components. Other details on the fit are consistent with the description in Sect.\thinspace\ref{sec:spectralanalysis}. The resulting best-fit normalizations per unit area of \texttt{apec}$_\mathrm{\texttt{LHB}}$, \texttt{apec}$_\mathrm{\texttt{MWH}}$, and \texttt{powerlaw} are displayed in \cref{fig:cxbnorms}. Thereafter, we used our CXB model to model the emission from the Antlia~SNR. The \texttt{nei} model consists of all four \texttt{apec} parameters and an additional parameter for the ionization time scale ($\tau$). We fixed their values to the best-fit values (\cref{tab:cxb parameters}) obtained from an extended region toward the south of Abell 1060 by \citet{2026A&A...708A.198K}. This region includes the majority of the foreground emission from the Antlia SNR between \rtwo\;and $\approx$4\rtwo\;in region B. Subsequently, we fixed the 
CXB normalizations of the rest of the CXB components to the best-fit values obtained from our initial fit and freed the \texttt{nei} normalization to fit the CXB spectra. We report a \mbox{c-stat/d.o.f. $\approx1$} for all the background boxes, and the obtained best-fit \texttt{nei} normalizations per unit area are displayed in \cref{fig:cxbnorms} (bottom right). \\
\indent
From \cref{fig:cxbnorms}, we observe the global trends in the CXB normalizations per unit area of all the components. The 0.1\thinspace\unit{keV} LHB and 0.25\thinspace\unit{keV} MWH components show minimal variations between the northwest, west, and southwest background boxes. Moreover, the northern box exhibits the lowest normalizations on average for LHB and MWH by 20.74\% and 28.3\%, respectively, as compared to the aforementioned boxes. On the eastern side, we see increments of 30.15\% and 13.5\%, respectively, in the LHB and MWH normalizations between the northeastern and the southern box, where both the normalizations are the highest. This result is consistent with the observed prominence of the foreground emission from region B and the extended foreground bubble from region A in the 0.2--0.8\thinspace\unit{keV} band, as observed in \cref{fig:wfrgb}. As for the harder background emission from the unresolved AGN, the highest and lowest normalizations between all the boxes vary only by a factor of 1.24. Furthermore, we note that the 0.18\thinspace\unit{keV} \texttt{nei} component shows the highest EM in the south and southeast, and declines azimuthally clockwise. Overall, the combined EM of the CXB is the highest in the south and gradually declines toward the northwest.     \\

\begin{table}[h]
   \centering
   \caption{Spectral parameters from the CXB analysis of the northwestern background box (``NW'' in Fig.\thinspace\ref{fig:nhtotmap}).}
   \label{tab:cxb parameters}
   \begin{tabular}{@{}cccc@{}}
   \toprule
   Component  & Parameter & Value & Status    \\ \midrule
   \texttt{apec}$_\mathrm{\texttt{LHB}}$&Norm [cm$^{-5}$arcmin$^{-2}$] & $2.55\substack{+0.10\\-0.13}\times10^{-6}$&Free\\
   &$k_\mathrm{B}T$ [keV]&0.1 &Fixed\\
   &Abundance $[Z_\mathrm{\odot}]$&1 &Fixed\\
   &Redshift&0&Fixed\\ \cmidrule{1-4}
   \texttt{apec}$_\mathrm{\texttt{MWH}}$&Norm [cm$^{-5}$arcmin$^{-2}$] & $6.55\substack{+0.37\\-0.28}\times10^{-7}$&Free\\
   &$k_\mathrm{B}T$ [keV]&0.25 &Fixed\\
   &Abundance $[Z_\mathrm{\odot}]$& 1&Fixed\\
   &Redshift&0&Fixed\\  \cmidrule{1-4}
   \texttt{nei}&Norm [cm$^{-5}$arcmin$^{-2}$] & $5.30\substack{+1.33\\-2.59}\times 10^{-8}$&Free\\
   &$k_\mathrm{B}T$ [keV]&0.18 &Fixed\\
   &$\tau$ [s cm$^{-3}$]&$2.20\times 10^{10}$&Fixed\\
   &Abundance $[Z_\mathrm{\odot}]$&1 &Fixed\\
   &Redshift&0&Fixed\\  \cmidrule{1-4}
   \texttt{powerlaw}&Norm$^{(a)}$&$4.82\substack{+0.36\\-0.26}\times10^{-7}$&Free\\
   &$\Gamma$&1.46&Fixed\\
   &Redshift&0&Fixed\\  \cmidrule{1-4}
   \texttt{TBabs}&$N_\mathrm{H}\;[\mathrm{cm}^{-2}]$&$5.59\times 10^{20}$&Fixed\\
   \bottomrule
   \end{tabular}
   \tablefoot{
      \tablefoottext{a}{Units are photons\thinspace\unit{keV}$^{-1}$ cm$^{-2}$ s$^{-1}$ arcmin$^{-2}$.}\\
      The ``Status'' column indicates whether the parameter was free or fixed during the ICM spectral analysis with the full model (Eq.\thinspace\ref{eq:clustermodel}).
   }
\end{table}
\indent
Therefore, we shortlisted the northwestern, western, and southwestern boxes to be a good representative of the local CXB within \rtwo\;because of the comparatively low EM from the Antlia SNR and the stable LHB and MWH EMs. Among these, the normalizations of the northwestern box are the closest to the median values across all regions, and its median $N_\mathrm{H,tot}$ is consistent with the median $N_\mathrm{H,tot}$ within \rtwo\;(\cref{fig:cxbnorms}). We further estimated the best-fit temperatures of the LHB and MWH components in this box by freeing their temperature parameters and fixing the \texttt{nei} parameters to the best-fit values in \cref{tab:cxb parameters}. We observed only a minor difference in the LHB component of +0.03\thinspace\unit{keV}, which produced a negligible median difference of 0.5\% in the temperature profile when used as the LHB temperature during the ICM analysis. Therefore, we used the default CXB parameters from \cref{tab:cxb parameters} for our ICM analysis.

\section{$N_\text{H,tot}$ correction simulations}
\begin{table}[h]
\label{tab:sim_params}
\centering
\caption{Parameter values for Eq. \ref{eq:sim_model}.}
\begin{tabular}{c c}
\toprule
Parameters& Value \\ \midrule
\texttt{apec}$_\mathrm{\texttt{LHB}}$ $k_\mathrm{B}T$ [keV] & 0.099 \\
\texttt{apec}$_\mathrm{\texttt{LHB}}$ Norm [cm$^{-5}$ deg$^{-2}$] & 0.0019 \\
\texttt{apec}$_\mathrm{\texttt{MWH}}$ $k_\mathrm{B}T$ [keV] & 0.225 \\
\texttt{apec}$_\mathrm{\texttt{LHB}}$ Norm [cm$^{-5}$ deg$^{-2}$] & 0.0041 \\
\texttt{powerlaw} $\Gamma$ & 1.4 \\
\texttt{powerlaw} Norm [photons keV$^{-1}$ cm$^{-2}$ s$^{-1}$ deg$^{-2}$] & 0.0036 \\
\hline
\end{tabular}
\end{table}

\section{Normalization, temperature, and metallicity profiles}
\begin{table*}[bp]
   \centering
   \begin{threeparttable}[b]
   \caption{Best-fit \texttt{apec}$_\mathrm{\texttt{ICM}}$ parameters obtained from seven radial annuli in the radial range $0\leq R \leq$ \rtwo.}
   \label{tab:tempprof}
   \begin{tabular}{@{}cccccc@{}}
   \toprule
   Radial bin  & Median $N_{\text{H,tot}}$&Normalization & $k_{\text{B}}T$ & Metallicity & c-stat/d.o.f.  \\ 
   $\left[\text{arcmin}\right]$  & $\left[\times 10^{20}\text{cm}^{-2}\right]$&$\left[\text{cm}^{-5}\;\text{arcmin}^{-2}\right]$ & $\left[\text{keV}\right]$ & $\left[Z_{\odot}\right]$ &   \\\midrule
   $0.00\leq R\leq 3.30$ & 6.34&$4.88\substack{+0.19 \\ -0.23}\times 10^{-4}$&$2.72\substack{+0.22 \\-0.20}$&$0.46\substack{+0.16 \\ -0.14}$ &6737.97/6981\\
   $3.30\leq R\leq 7.85$ &6.29 &$1.74\substack{+0.08 \\ -0.08}\times 10^{-4}$ &$2.78\substack{+0.24 \\-0.20}$ & $0.40\substack{+0.14\\ -0.13}$ & 7352.61/7422\\
   $7.85\leq R\leq 13.20$& 6.24&$7.07\substack{+0.33 \\ -0.33}\times 10^{-5}$ &$3.00\substack{+0.34 \\-0.28}$ & $0.10\substack{+0.15\\ -0.10}$ & 7580.38/7664\\
   $13.20\leq R\leq 19.29$& 6.13&$2.83\substack{+0.17 \\ -0.17}\times 10^{-5}$ &$2.08\substack{+0.27 \\-0.24}$ & $0.22\substack{+0.12\\ -0.10}$ & 7535.80/7908\\
   $19.29\leq R\leq 27.62$& 6.00&$1.06\substack{+0.01 \\ -0.01}\times 10^{-5}$ &$1.77\substack{+0.32 \\-0.23}$ & $0.22\substack{+0.14\\ -0.12}$ & 8028.02/8534 \\
   $27.62\leq R\leq R_\mathrm{500}$ & 5.76 &$3.28\substack{+0.36 \\ -0.55}\times 10^{-6}$ &$1
   .45\substack{+0.20 \\-0.10}$ & $0.31\substack{+0.17\\ -0.13}$ & 9861.76/10212 \\ 
   $R_\mathrm{500}\leq R \leq R_\mathrm{200}$& 5.69&$3.19\substack{+0.08\\-0.09}\times 10^{-6}$&$1.51\substack{+0.06 \\ -0.07}$&$0.10-0.50$& 11028.15/10842\\ \midrule
   $0.2R_{\text{500}}\leq R \leq0.5R_{\text{500}}$ & 6.12&$3.04\substack{+0.12 \\ -0.12}\times 10^{-5}$ &$2.51\substack{+0.21 \\-0.21}$ & $0.19\substack{+0.10\\ -0.10}$ & 8550.49/8993  \\
   \bottomrule
   \end{tabular}
   \tablefoot{The second column lists the median $N_\mathrm{H,tot}$ per bin. For the $R_\mathrm{500}$--$R_\mathrm{200}$ bin, the mean c-stat across all fits is listed.}
   \end{threeparttable}
\end{table*}
In \cref{tab:tempprof} we display the best-fit parameters from the ICM spectral analysis.
\end{appendix}

\end{document}